%% file: main_nmi.tex
\documentclass[journal=jcim,manuscript=article]{achemso}

\usepackage{times}
\usepackage{soul}
\usepackage{url}
\usepackage[hidelinks]{hyperref}
\DeclareUnicodeCharacter{0307}{\.}
\DeclareUnicodeCharacter{0301}{\'{e}}
\usepackage[utf8]{inputenc}
\usepackage[T1]{fontenc}
\usepackage{graphicx}
\usepackage{amsmath,amssymb,amsfonts}
\usepackage{amsthm}
\usepackage{booktabs}
\usepackage{enumitem}
\usepackage[dvipsnames]{xcolor}
\usepackage{xspace}
\usepackage[flushleft]{threeparttable}
\usepackage{subcaption}
\usepackage{caption, mathtools}
\usepackage{comment}
\usepackage{algorithm}
\usepackage{algorithmic}
\usepackage{multirow}
\usepackage{xr}
\include{define}

\SectionNumbersOn
\mciteErrorOnUnknownfalse
\title{\method: Yield Prediction via Local-to-global Reaction Representation Learning and Interaction Modeling}

\author{Xiao Hu}
\affiliation{Computer Science and Engineering, The Ohio State University,\\Columbus, OH, USA, 43210}
\author{Ziqi Chen}
\affiliation{Computer Science and Engineering, The Ohio State University,\\Columbus, OH, USA, 43210}
\author{Bo Peng}
\affiliation{Computer Science and Engineering, The Ohio State University,\\Columbus, OH, USA, 43210}
\author{Daniel Adu-Ampratwum}
\affiliation{Division of Medicinal Chemistry and Pharmacognosy, College of Pharmacy, The Ohio State University,\\Columbus, OH, USA, 43210}
\author{Xia Ning}
\affiliation{Computer Science and Engineering, The Ohio State University,\\Columbus, OH, USA, 43210}
\alsoaffiliation{Division of Medicinal Chemistry and Pharmacognosy, College of Pharmacy, The Ohio State University,\\Columbus, OH, USA, 43210}
\alsoaffiliation{Translational Data Analytics Institute, The Ohio State University,\\Columbus, OH, USA, 43210}
\alsoaffiliation{Biomedical Informatics, The Ohio State University,\\Columbus, OH, USA, 43210}
\email{ning.104@osu.edu}

\begin{document}

\maketitle

\begin{abstract}
Accurate prediction of chemical reaction yields is crucial for optimizing organic synthesis, potentially reducing time and resources spent on experimentation. With the rise of artificial intelligence (AI), there is growing interest in leveraging AI-based methods to accelerate yield predictions without conducting in vitro experiments.
We present \method, an innovative graph transformer-based framework designed for predicting chemical reaction yields. A key feature of \method is its integration of a cross-attention mechanism that focuses on the interplay between reagents and reaction centers. This design reflects a fundamental principle in chemical reactions: the crucial role of reagents in influencing bond-breaking and formation processes, which ultimately affect reaction yields.
\method also implements a local-to-global reaction representation learning strategy. This approach initially captures detailed molecule-level information and then models and aggregates intermolecular interactions. Through this hierarchical process, \method effectively captures how different molecular fragments contribute to and influence the overall reaction yield.
%
\method shows superior performance in our experiments, especially for medium to high-yielding reactions, proving its reliability as a predictor.
The framework's sophisticated modeling of reactant-reagent interactions and precise capture of molecular fragment contributions
make it a valuable tool for reaction planning and optimization in chemical synthesis.
The data and codes of \method are accessible through \href{https://github.com/ninglab/Yield_log_RRIM}{https://github.com/ninglab/Yield\_log\_RRIM}. 
\end{abstract}

\section{Introduction}
%
Chemical yield prediction is crucial for optimizing organic synthesis, offering chemists an efficient tool to identify high-yielding reactions while reducing time and resource expenditure~\cite{shields2021bayesian}. 
Traditionally, chemists have relied on expertise and systematic experimentation to optimize reactions~\cite{reizman2012automated}. While foundational, these methods can become resource-intensive when scaling up~\cite{sigman2016development}. 
Consequently, there is an increasing interest in developing artificial intelligence (AI)-based methods~\cite{schwaller2021prediction, probst2022reaction, lu2022unified, chen2021sequence, li2021mol, saebi2023use,li2023reaction}. These AI-based methods allow chemists to accelerate precise yield prediction without doing in vitro experiments, potentially enhancing the efficiency of organic synthesis optimization.
Despite the importance of the task, AI-based computational methods have received comparatively little attention in yield prediction compared to other chemistry-related tasks (e.g. forward prediction~\cite{schwaller2019molecular, coley2019graph}, retrosynthesis~\cite{liu2017retrosynthetic,chen2023g}).
We aim to bridge the gap and introduce novel and effective AI methods for yield prediction. 

Early AI-based methods focused on identifying effective chemical knowledge-based reaction descriptors~\cite{kariofillis2022using,yada2018machine} and employing traditional machine learning models~\cite{cortes1995support,breiman2001random} over such descriptors for chemical yield prediction. However, these methods often produce unsatisfactory results, suggesting the limitation of the chemical knowledge-based descriptors, as well as the companion traditional machine learning models.
%
%
The advent of language models~\cite{raffel2020exploring, devlin2018bert} has enabled sequence-based approaches for chemistry-related tasks~\cite{schwaller2021prediction, probst2022reaction, lu2022unified, chen2021sequence, li2021mol}. These models are typically pre-trained on large molecular datasets~\cite{kim2019pubchem} using SMILES~\cite{weininger1988smiles} representations and then fine-tuned on specific datasets for yield prediction with the entire reaction's SMILES string as input. However, this pre-training and fine-tuning framework may not be optimal for yield prediction~\cite{schwaller2021prediction, lu2022unified}, as it lacks features that account for unique characteristics of the task, such as explicit modeling of reactant-reagent interactions.
Moreover, these models, using the entire reaction as input, tend to overlook the contributions of small yet influential molecular fragments~\cite{neel2016enantiodivergent}, as their attention mechanisms may not be sensitive enough to focus on these critical elements. 
Additionally, building such pre-trained foundation models is resource-intensive.
In contrast to the sequence-based models, graph neural networks (GNNs)
have recently been employed to represent molecules and reactions as graphs, learning molecular structural information for yield prediction~\cite{saebi2023use,li2023reaction}. This approach allows for a more intuitive representation of molecular structure compared to sequence-based models. However, most GNN-based methods still lack effective modeling of molecular interactions.
This limitation is particularly significant in yield prediction, as the interactions between reactants and reagents, like catalysts, can substantially impact reaction outcomes~\cite{carlson2005design,coley2018machine}.

To address these challenges, we introduce \method: a graph transformer-based \underline{lo}cal-to-\underline{g}lobal \underline{r}eaction \underline{r}epresentation learning and \underline{i}nteraction \underline{m}odeling for yield prediction.
\method incorporates a cross-attention mechanism between the reagents and reaction center atoms to simulate a principle of chemical reactions: reagents have a huge impact on the bond-breaking and formation of the reaction, thus affecting the yield changes. This design more effectively captures the interactions between molecules (reactants and reagents), thereby improving the prediction accuracy.
Additionally, \method employs a local-to-global graph transformer-based reaction representation learning process, which first learns representations at the molecule level for each component individually and then models their interactions. This information is then aggregated, ensuring a more balanced attention mechanism that considers molecules of all sizes, preventing small fragments from being overlooked in the whole reaction for yield prediction.
%

%
Performance evaluation on the commonly investigated datasets~\cite{ahneman2018predicting,lu2022unified,jiang2021smiles} demonstrates \method's superior prediction accuracy, particularly for medium to high-yielding reactions. 
This suggests its potential for enhancing reaction yield optimization accuracy in practical synthetic chemistry.
Our analyses further reveal \method's effectiveness in capturing complex molecular (reactant-reagent) interactions and accurately assessing small molecular fragments' contributions to yield. 
These capabilities highlight \method's potential for optimizing synthetic routes through informed modifications of reactants and reagents, providing chemists with a sophisticated instrument for reaction design and optimization. 
%

%

\section{Related Work}
Reaction yield prediction has evolved primarily through three types of approaches, each addressing the challenges of representing complex molecular structures and modeling their interactions in different ways.
The approaches started with traditional machine learning models based on chemical knowledge-based descriptors. 
Next, sequence-based models were developed, representing each molecule as a SMILES string. These models are typically pre-trained on large molecule datasets to learn general molecule representations and then fine-tuned specifically for yield prediction tasks.
Most recently, graph-based models have emerged as a powerful tool for learning molecular structures, treating molecules as graphs, and aggregating molecular information for prediction. 

\subsection{Traditional Machine Learning Models}
Early approaches to yield prediction utilized traditional machine learning models, such as random forest (RF)~\cite{breiman2001random} and support vector machine (SVM)~\cite{cortes1995support}, to predict yields. These models relied on chemical knowledge-based descriptors to depict the molecule properties, which include density functional theory calculations~\cite{kariofillis2022using,yada2018machine}, one-hot encoding ~\cite{chuang2018comment}, and fingerprint features ~\cite{sandfort2020structure}.
These methods were primarily evaluated on reaction datasets containing a single reaction class~\cite{ahneman2018predicting, perera2018platform}.
However, they often demonstrated unsatisfactory performance~\cite{kariofillis2022using,yada2018machine,chuang2018comment,sandfort2020structure}. This highlighted two main limitations. First, the modeling ability of traditional machine learning methods is insufficient for this complex problem. Second, relying solely on pre-defined chemical descriptors for constructing reaction representations is inadequate.
The suboptimal results obtained from these methods suggest that more sophisticated and effective approaches are needed to capture the complex information between molecular structures and reaction yields. 
\subsection{Sequence-based Models}
Transformer-based models have recently gained prominence in chemical tasks~\cite{schwaller2021prediction, probst2022reaction, lu2022unified, chen2021sequence, li2021mol}.
These models are typically pre-trained on large molecular datasets represented by SMILES strings, learning general molecular representations. They are then fine-tuned on specific datasets containing yield information for the prediction. During fine-tuning, the models learn to process the SMILES string of the entire reaction as input, enabling them to capture relationships between all reaction components. 
%
For example, Schwaller \etal introduced \yieldbert~\cite{schwaller2021prediction}, which employs the SMILES string of a whole reaction as input to a BERT-based yield predictor~\cite{devlin2018bert}. This BERT-based yield predictor is obtained from fine-tuning a yield regression head layer on a reaction encoder~\cite{schwaller2021mapping}. 
Similarly, Lu and Zhang developed \chemt~\cite{lu2022unified}, utilizing the Text-to-Text Transfer Transformer (T5) model~\cite{raffel2020exploring}. \chemt, pre-trained on the PubChem dataset~\cite{kim2019pubchem}, is designed for multiple reaction prediction tasks (e.g., product prediction, retrosynthesis) and employs a fine-tuned regression head for yield prediction purposes.
The sophisticated sequence modeling techniques enable these methods to learn more informative reaction representation than handcrafted chemical knowledge-based descriptors by capturing contextual information embedded in the SMILES string of the entire reactions. 
Consequently, they demonstrate commendable prediction performance on datasets containing a single reaction class. 

However, the efficacy diminishes when testing on datasets with a wide variety of reaction types and diverse substances, such as the US Patent database (USPTO)~\cite{Lowe2017}. 
Additionally, treating the whole reaction as input makes it challenging for the sequence-based models to distinguish the effects of different components in a reaction, as reactants and reagents have distinct impacts on yield.
Also, small modifications in the molecules, even those involving only a few fragments (atoms, functional groups, or small-size molecules), can significantly affect reaction outcomes~\cite{neel2016enantiodivergent}. When sequence-based models treat the entire reactions as inputs, they tend to overlook the contributions of those small yet influential fragments. This occurs because the attention mechanisms used in these sequence-based models may not be sufficiently sensitive to those critical fragments, potentially leading to inaccurate predictions.

To address these challenges, we propose to apply a local-to-global learning process to ensure equal attention is allocated to molecules of varying sizes. The local-to-global learning process treats each reactant, reagent, and product separately before interacting and aggregating their information, intuitively depicting the role of different components in the reaction. This prevents the model from ignoring the impact of small fragments. Our experiment and analysis demonstrate the effectiveness of our modeling design.
\subsection{Graph-based Models}
%
Recent advancements have established graph neural networks (GNNs) as powerful tools for analyzing molecules and predicting reaction yields~\cite{yang2019analyzing, jo2020message, tang2023application, yarish2023advancing, saebi2023use,li2023reaction}. These approaches represent chemical structures as graphs, using GNNs to learn structural information and typically employing multilayer perceptrons (MLPs) to predict yields after aggregating molecular information into vector representations.
Saebi \etal developed \yieldgnn~\cite{saebi2023use}, which uses Weisfeiler-Lehman networks (WLNs)~\cite{lei2017deriving} to aggregate atom and bond features over their neighborhood and finally obtain the high-order structural information. These learned structural features and the selected chemical knowledge-based reaction descriptors are then combined to predict the reaction yield through a linear layer. 
Their results highlight the importance of learned molecular structural features over the chemical descriptors.
Yarish \etal introduced \rdmpnn~\cite{yarish2023advancing}, 
which first uses directed message passing networks (D-MPNN)~\cite{yang2019analyzing} to generate atom and bond embeddings from reactant and product graphs. Then, it creates the chemical transformation encoding according to the atom and bond mapping between the reactants and the products, which is combined with pre-computed molecular descriptors to predict the yield.
Li \etal proposed \semg~\cite{li2023reaction}, which similarly employs a GNN to update atom features and obtain molecule representations. Then, it applies an attention mechanism based on all involved components to model the molecular interplays and derive the reaction representation for prediction. 

While these graph-based methods demonstrate satisfactory performance on datasets of a single reaction class, they have not been extensively tested on challenging datasets like USPTO. 
Furthermore, these approaches exhibit certain limitations in molecular interaction design. \rdmpnn and \yieldgnn lack explicit modeling of interactions among reactants and reagents, while \semg does not effectively capture the full complexity of molecular interactions.

To address these limitations and better enable the model to learn the interactions between reactants and reagents, we propose to explicitly characterize the function of reagents on the reaction center.
This approach uses a cross-attention mechanism~\cite{vaswani2017attention} to capture the complex interplay between different reaction components (reactants and reagents) more effectively, potentially leading to improved yield predictions. Our experiments and analysis demonstrate that this design improves the effectiveness of molecular interaction modeling.

\section{Materials}

\subsection{Datasets}
\subsubsection{USPTO500MT Dataset} 
USPTO500MT is derived from USTPO-TPL~\cite{schwaller2021mapping} by the authors of \chemt ~\cite{lu2022unified}. USTPO-TPL comprises 445,000 reactions, with yield reported, partitioned into 1,000 strongly imbalanced reaction types. 
USPTO500MT is obtained by extracting the top 500 most frequently occurring reaction types from USPTO-TPL. It consists of 116,360 reactions for training, 12,937 reactions for validation, and 14,238 reactions for testing purposes. 
The reactants, reagents, and products are encoded as SMILES strings. The yield distribution is summarized in Figure~\ref{fig:uspto_yield_dist} and the entire dataset is skewed towards high-yielding reactions.
Within the USPTO500MT dataset, approximately 95.5\% of the reactions (129,437) are unique. Additionally, about 3.7\% of the products (4,949) are documented with two distinct synthesized processes. Only a small fraction (0.1\%) of products are synthesized through over five different processes.
Moreover, the number and the function of reagents are varying among each reaction.
These showcase the diversity and complexity of the reactions within the dataset. 
\subsubsection{Buchwald–Hartwig Amination Reaction Dataset} 
\begin{figure*}[h!]
\begin{center}
\begin{subfigure}[t]{0.4875\linewidth}
    \centering
    \includegraphics[width=\linewidth]{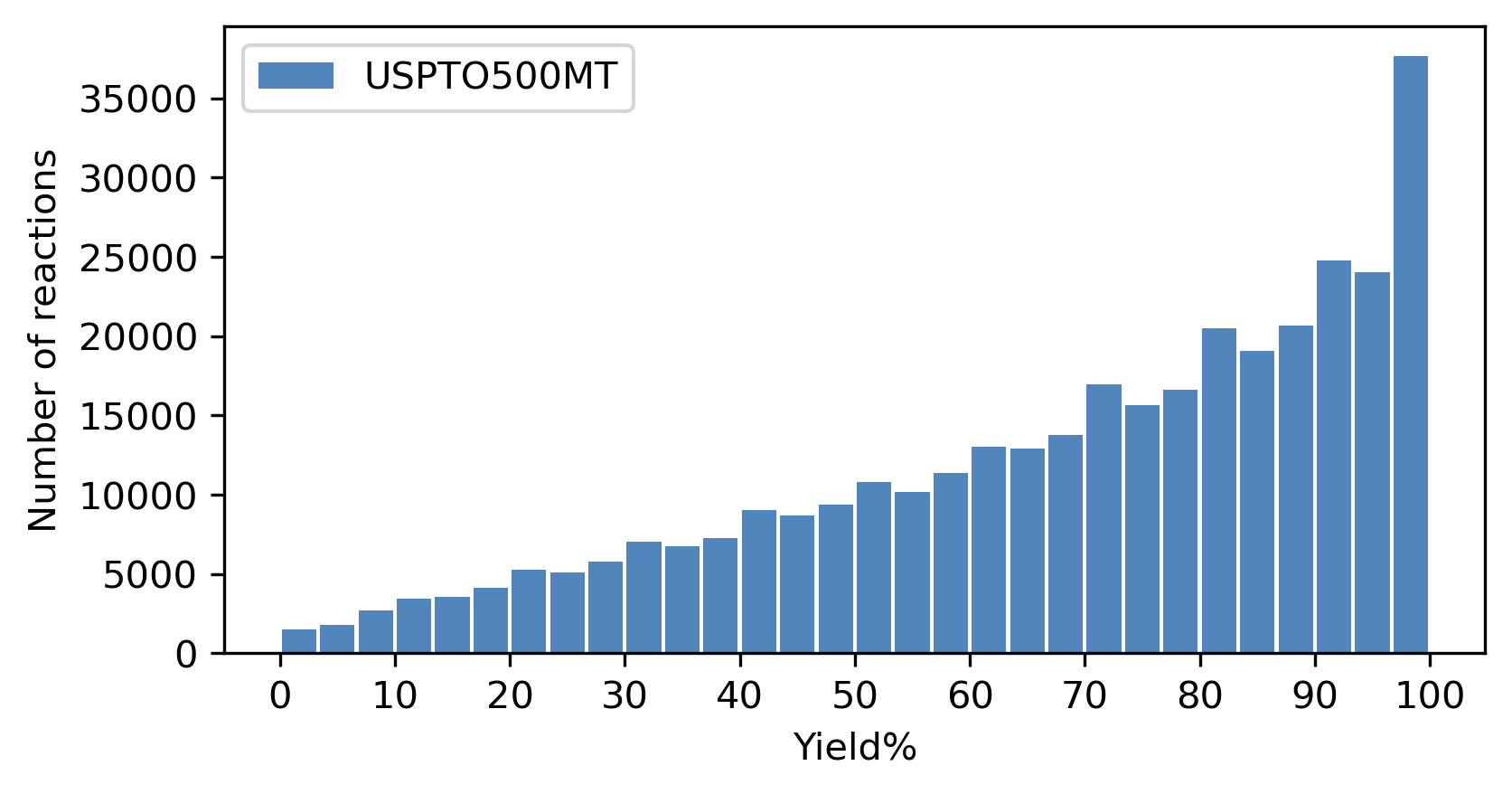}
    \caption[b]{USPTO500MT reactions yield distribution}
    \label{fig:uspto_yield_dist}
 \end{subfigure}
\begin{subfigure}[t]{0.47\linewidth}
    \centering
    \includegraphics[width=\linewidth]{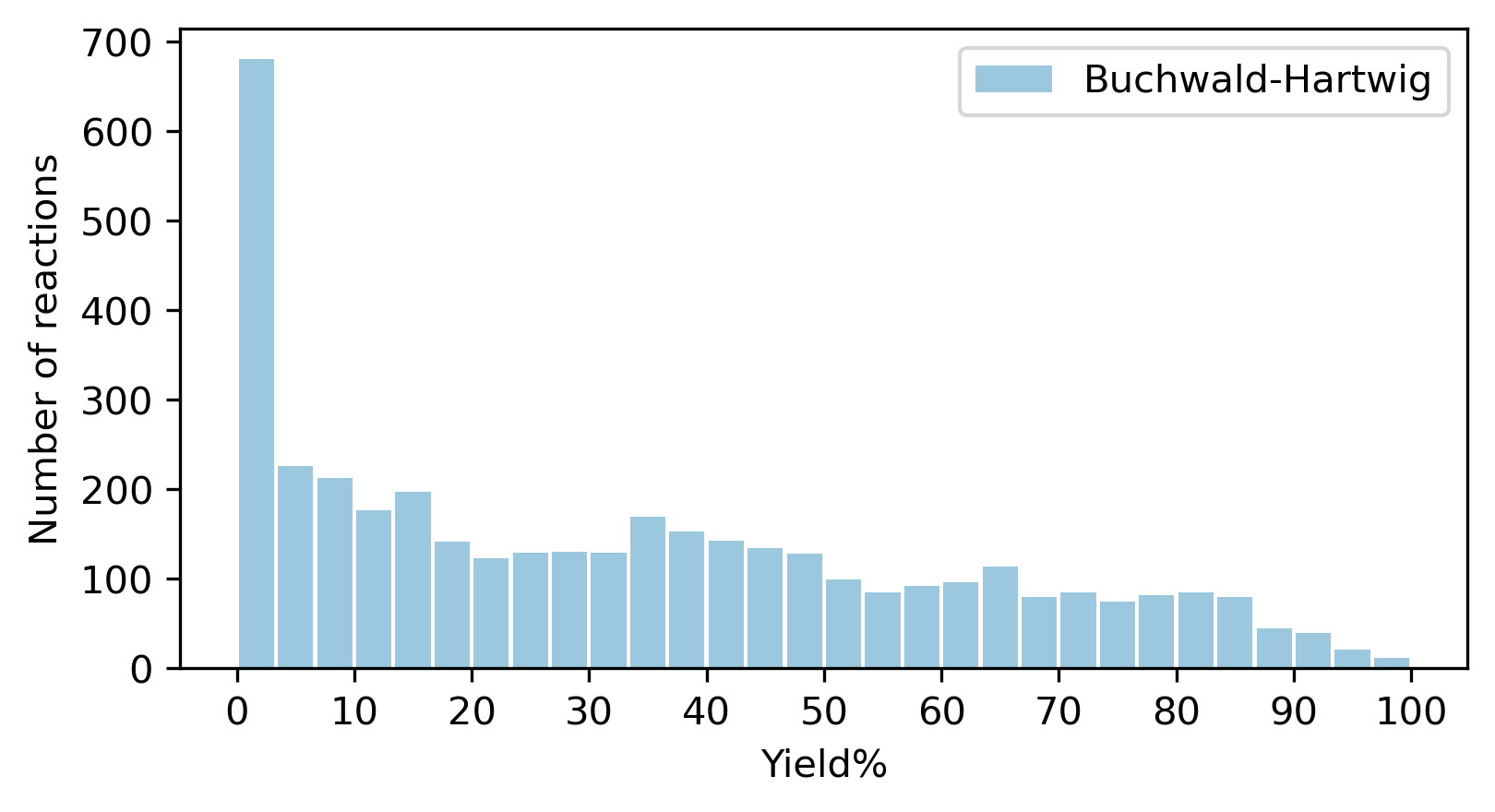}
    \caption[a]{Buchwald–Hartwig reactions yield distribution}
    \label{fig:bh_yield_dist}
\end{subfigure}\hfil

\caption{Overview reactions yield distributions of the two datasets}
\label{fig:yield_dists}
\end{center}
\end{figure*}

The Buchwald-Hartwig dataset, constructed by Ahneman \etal~ \cite{ahneman2018predicting}, has become a benchmark for assessing the performance of yield prediction models. This dataset comprises 3,955 palladium-catalyzed C-N cross-coupling reactions, with yields obtained through high-throughput experimentation (HTE).
The dataset encodes information on reactants, reagents, and products as SMILES strings. It includes 15 distinct aryl halides paired with a single amine as reactants. These reactant pairs undergo experimentation with 3 different bases, 4 Buchwald ligands, and 22 isoxazole additives, resulting in 5 different products. 
The yield distribution, illustrated in Figure~\ref{fig:bh_yield_dist}, reveals a notable skew due to a substantial proportion of non-yielding reactions.

In comparison to broader datasets such as USPTO500MT, the Buchwald-Hartwig dataset is limited to a single reaction type and features a constrained set of reaction components. 
Moreover, reagent information is consistently organized, with each reaction entry containing ligand, base, and solvent information in a consistent order. 
While this structured format may facilitate easier predictive model learning, it potentially misrepresents real-world scenarios where the data is often comprehensive and less organized. 
This underscores the limitation in this dataset's ability to reflect the complexity and variability of practical chemical information, despite its value as a benchmark for yield prediction models.

\subsection{Training data generation}
\subsubsection{Basic atom features}
We follow Maziarka \etal ~\cite{maziarka2020molecule} and employ the open-source RDKit toolkit to extract the basic chemical features for atoms in molecules represented by SMILES strings. The basic atom features utilized in \method are delineated in Table~\ref{tbl:Atom_features}. These features describe the basic chemical properties and environment, serving as the input of \methodwo.
\input{tables/Atom_features}

\subsubsection{Learned atom representations from pre-trained models} 
To investigate the impact of atom features chosen on \method, we employ two approaches: one using the basic atom features directly, and another using learned atom representations derived from a pre-trained model MAT by Maziarka \etal~\cite{maziarka2020molecule}. 
We name the \method trained on basic atom features as \methodwo, and the version trained on learned atom representations as \methodw.
The pre-trained model MAT takes the basic atom features as input and utilizes node-level self-supervised learning ~\cite{hu2019strategies} on a subset of 2 million molecules from the Zinc15 dataset ~\cite{sterling2015zinc} for molecule representation learning. 
These learned atom representations are then input for \methodw, potentially capturing more complex atomic relations and information.
The hyperparameters of the pre-trained model are delineated in Table~\ref{tbl:pretrained_MAT_hyperparameters} and remain consistent across all experiments.
\subsubsection{Reaction center identification}
Identifying reaction centers is crucial for \method as it allows us to pinpoint the specific atoms involved in the chemical transformation.
We follow GraphRetro's~\cite{somnath2021learning} approach to identify these reaction center atoms by comparing the changed bonds between the mapped reactant and product molecules. 
In \method, we model the interactions between these reaction centers and reagents, which enables us to more effectively capture the key information (reagents have an impact on bond-breaking and formation) that influences the reaction yield, potentially improving the accuracy of predictions.
\subsection{Experimental setting} 
%
For the USPTO500MT dataset, we adopt the training, validation, and testing split used by \chemt. 
We adhere to the data-splitting protocol for the Buchwald-Hartwig dataset as \yieldgnn, using 10-fold 70/30 random train/test splits. We further allocate 10\% of the training data for validation. After determining the optimal hyperparameters using three data splits, we apply the model across all ten data splits and compare its performance against other baselines.
In addition, we exclude reactions that cannot be processed by the reaction center identification method. 
The reaction center identification process ensures that all reactions in the dataset have well-defined reaction centers and identifiable mechanistic pathways, which is critical for accurate modeling of reaction mechanisms and yield predictions. 
While this process does not filter out any reactions in the Buchwald–Hartwig dataset, it results in 78,201 reactions filtered out for training, 8,716 for validation, and 9,497 for testing in USPTO500MT. This curation enhances the overall integrity of USPTO500MT, allowing for more precise reactions to be considered.
All performance comparisons are conducted on these curated datasets to maintain consistency in our evaluations.
\subsection{Model evaluation} 
The range of ground-truth yield is between 0-1 (0\%-100\%). We use mean absolute error (MAE) and root mean squared error (RMSE) to evaluate the prediction performance. Their calculations are given by the following equations:
\begin{equation}  
\label{Equation:mse}
\text { MAE }=\frac{\sum_{i=1}^N\left|y_i-\hat{y}_i\right|}{N},
\end{equation}

\begin{equation}
\label{Equation:rmse}
\text { RMSE }=\sqrt{\frac{\sum_{i=1}^N\left(y_i-\hat{y}_i\right)^2}{N}},
\end{equation}
where $\hat{y}_i$ is the predicted yield, $y_i$ is the ground-truth yield, and $N$ is the number of samples. The smaller the MAE and RMSE are, the more accurate the yield predictor model is.
Previous methods~\cite{ahneman2018predicting,lu2022unified} use the coefficient of determination ($R^2$) to evaluate the goodness of fit of the regression model, which is defined as follows:
\begin{equation}
\label{Equation:r2}
\text { $R^2$ }=1-\frac{\sum_{i=1}^N\left(y_i-\hat{y}_i\right)^2}{\sum_{i=1}^N\left(y_i-\bar{y}\right)^2},
\end{equation}
where $\bar{y}$ is the mean of $N$ ground-truth yields and a larger value of $R^2$ implies a better goodness of fit of the models.
However, $R^2$ is not an ideal metric to evaluate the accuracy and relationship, as it has several limitations~\cite{balaji2023traumatic}. One significant issue is that $R^2$ can be heavily influenced by outliers, potentially giving a distorted view of the model's overall fit. This sensitivity means that a few extreme error predictions can lead to a very low $R^2$, even the majority of predictions are accurate. 
Therefore, it is challenging to draw definitive conclusions from $R^2$, especially when it is low. While we still present the results in $R^2$ in line with the literature, the evaluation is primarily via MAE and RMSE.

\section{Method}
Our method, \method, is a novel local-to-global graph-transformer-based reaction representation learning and molecular interaction modeling framework.
It employs a local-to-global learning process for reaction representation learning, beginning with molecule (reactants, reagents, and product) representation learning. It subsequently models the molecule interactions (between reactants and reagents) and ultimately represents the entire reaction. 
\method further uses the reaction representation to predict yield. 

\begin{figure*}[h!]
\begin{center}
{\includegraphics[scale=0.48]{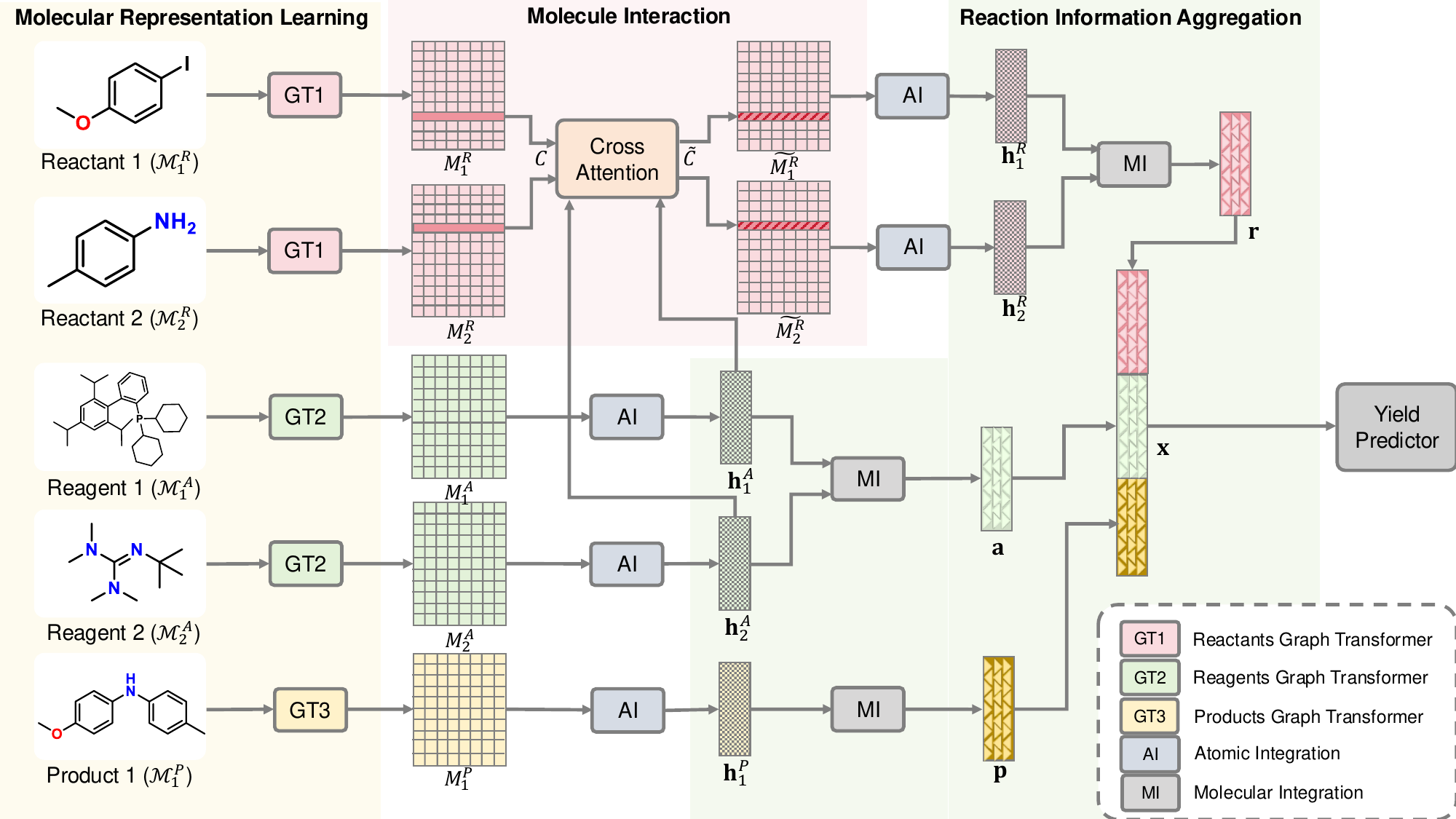}}
\caption{Pipeline of \method}
\label{pipeline}
\end{center}
\end{figure*}

Specifically, \method consists of the following three modules:
 \textbf{(1)} Molecule Representation Learning (\MRL) module: 
which uses graph transformers~\cite{maziarka2020molecule} with multi-head self-attention layers to encode molecular structural information into atom embeddings, and then aggregate atom embeddings 
into molecule embeddings through Atomic Integration (\AI).
 \textbf{(2)} Molecule Interaction (\MIT) module: which learns the interactions between reactants and reagents through the cross-attention mechanism, resulting in interaction-aware embeddings for reaction centers.
 %
 \textbf{(3)} Reaction Information Aggregation (\RIA) module: which employs Molecular Integration (\MI) to derive a comprehensive reaction representation from all involved molecules and their interaction representations.
 Finally, this reaction representation is utilized to predict the yield.
 An overview of \method is depicted in Figure~\ref{pipeline}.

\subsection{Notations} 

In a reaction $\mathcal{X}$, each reactant, reagent, and product is a molecule. We view each molecule $\mathcal{M}$ as a graph, with basic node (atom) features $I\in \mathbb{R}^{n\times {s}}$, graph adjacent matrix $J \in\{0,1\}^{n \times n}$,
and inter-atomic distance matrix $D \in \mathbb{R}^{n \times n}$, where $n$ is the number of atoms in the molecule and can be different for each molecule, $s$
is the dimension of basic atom features.
The reaction $\mathcal{X}$ is represented as $(\mathcal{R}, \mathcal{A}, \mathcal{P}, y)$, where $\mathcal{R}=\left\{\mathcal{M}^{R}_1, \ldots, \mathcal{M}^{R}_{n_r}\right\}$, $\mathcal{A}=\left\{\mathcal{M}^{A}_1, \ldots, \mathcal{M}^{A}_{n_a}\right\}$, and $\mathcal{P}=\left\{\mathcal{M}^{P}_{1}, \ldots, \mathcal{M}^{P}_{n_p}\right\}$ are the set of $n_r$ reactants, $n_a$ reagents and $n_p$ products in the reaction, and $y$ is the reaction yield. $n_r$, $n_a$, and $n_p$ can be different for each reaction.
%
Notably, we denote the reaction center atom embeddings in reactants as $C \in \mathbb{R}^{|C| \times d}$, where $|C|$ refers to the number of reaction center atoms.

In \MRL module, the atom embeddings of each molecule after the $l \in[1, n_l]$-th self-attention layer is denoted as ${M}(l)\in \mathbb{R}^{n\times {d}}$, 
where $n_l$ is the number of self-attention layers used in \MRL and $d$ is the model dimension. 

$\mathbf{h}_m \in \left\{\mathbf{h}^{R}_{1}, \ldots, \mathbf{h}^{R}_{n_r}, \mathbf{h}^{A}_{1}, \ldots, \mathbf{h}^{A}_{n_a}, \mathbf{h}^{P}_1, \ldots, \mathbf{h}^{P}_{n_p}\right\} $ is a $d$-dimension vector and denoted as the representation of each molecule in reactants, reagents, and products. The reactant, reagent, and product representations are named as $\mathbf{r}\in \mathbb{R}^{d}$, $\mathbf{a}\in \mathbb{R}^{d}$ and $\mathbf{p}\in \mathbb{R}^{d}$. 
%
The representation of the whole reaction is $\mathbf{x}$ 
and the predicted yield is denoted by $\hat{y}$. We summarize the key notations in Table~\ref{tbl:Notations}. 
%
We use uppercase letters to denote matrices, lowercase bold letters to denote row vectors, and lower-case non-bold letters to represent scalars.

\input{tables/Notations}


\subsection{Molecule Representation Learning (\MRL)} 
Given a reaction consisting of molecules represented in graphs, we first employ the Molecule Attention Transformer (MAT) ~\cite{maziarka2020molecule} to learn the molecule representations. 
MAT recursively propagates information between atoms to learn the structural information of molecules via multi-head molecule self-attention layers as follows:
\begin{equation}
\begin{aligned}
& Q_j^l=M(l-1) E_j^l, K_j^l=M(l-1) F_j^l, V_j^l=M(l-1) G_j^l,  \\
& \text{HEAD}_j^l=\left(\lambda_a \rho\left(\frac{Q_j^l\left(K_j^l\right)^T}{\sqrt{d}}\right)+\lambda_d \rho(D)+\lambda_g J\right) V_j^l,  \\
& {H}^l=\left[\text{HEAD}_1^l, \text{HEAD}_2^l, \ldots, \text{HEAD}_{n_h}^l\right] O^l, 
\end{aligned}
\end{equation}
where $M(l-1)$ is the atom embeddings from the $(l-1)$-th molecule self-attention layer and $M(0)=I$ is the input of first layer
; $Q_j^l$, $K_j^l$, and $V_j^l$ are the query, key, and value matrix derived from $M(l-1)$ with learnable parameters $E_j^l$, $F_j^l$ and $G_j^l \in \mathbb{R}^{d \times \frac{d}{n_h}}$;
$\lambda_a, \lambda_d, \lambda_g$ are the scalars to balance the importance of the self-attention, distance matrix, and adjacency matrix, and $\rho$ is the softmax function.
Each molecule attention layer has $n_h$ heads and $\text{HEAD}_j^{l}$ is the output from the $j$-th attention head; $O^l \in \mathbb{R}^{d\times {d}}$ is a learnable matrix to integrate the attention heads.
After each molecule self-attention layer, MAT includes a feed-forward layer to introduce non-linearity which is a fully connected network (FCN) described below:
\begin{equation}
{M}(l)=\sigma\left(H^l W^l+B^l\right),
\end{equation}
where $W^l \in \mathbb{R}^{d\times {d}}$ and $B^l \in \mathbb{R}^{n\times {d}}$ are learnable parameters, $\sigma(\cdot)$ is ReLU~\cite{nair2010rectified} activation function.
After $n_l$ molecule self-attention layers, the molecule's structural information is encoded into the atom embeddings $M(n_l)$. When no ambiguity arises, for simplicity, we eliminate $n_l$ for ${M}(n_l)$ and only use ${M}$ to represent atom embeddings of molecule $\mathcal{M}$ as the output of the last molecule self-attention layer. 

Compared to the original Transformer~\cite{vaswani2017attention}, MAT integrates the interactions among atoms, the geometric information in the molecule, and the topology of the molecule to better learn expressive atom embeddings, and captures the structural information of the molecule.
Given the atom embeddings $M$ learned from MAT, we utilize the Atomic Integration(\AI) module to aggregate atom embeddings and generate the molecule representation $\mathbf{h}_m$.
Particularly, \AI uses a gating mechanism to capture the importance of different atoms in the aggregation as follows:
\begin{equation}
\begin{aligned}
& \boldsymbol{\alpha}=M \mathbf{w}_1, \\
& \mathbf{h}_m=\sum_{k=1}^{n}\left[M\right]_k \times[\boldsymbol{\alpha}]_k,
\end{aligned}
\end{equation}
%
where $\mathbf{w}_1 \in R^{d}$ is a learnable vector and $\boldsymbol{\alpha}$ is the vector where each element represents the contribution of each atom embedding to the molecule representation. 
%

%
%
%
Additionally, in the \MRL module, \AI is only performed on reagents and products to get their molecule representations and is omitted for reactants. 
This is because the reaction center atom embeddings in the reactants will undergo further updates in the Molecule Interaction(\MIT) module. The reactant molecule representations will be obtained through \AI afterward.
%
%

\subsection{Molecule Interaction (\MIT)} 
Reagents, such as catalysts, significantly impact reaction yield by promoting or inhibiting bond breaking and formation. We explicitly model their function to the reaction center atoms to better capture the interaction between reactants and reagents.
Specifically, given the reaction center atom embeddings $C \in \mathbb{R}^{|C|\times d}$ in reactant molecules, and the reagent molecule representations $\mathbf{h}_i^{A} \in \left\{ \mathbf{h}^{A}_1, \ldots, \mathbf{h}^{A}_{n_a}\right\}$, we update the reaction center atom embeddings by applying a multi-head cross-attention mechanism, described as follows:
\begin{equation}
\begin{aligned}
& Q_j=C W_j^Q, K_j=H^{A} W_j^K, V_j=H^{A} W_j^V,  \\
& \text{HEAD}_j=\rho\left(\frac{Q_j\left(K_j\right)^T}{\sqrt{d}}\right) V_j,  \\
& H=\left[\text{HEAD}_1, \text{HEAD}_2, \ldots, \text{HEAD}_{n_h}\right] O, 
\end{aligned}
\end{equation}
where $Q_j$ is the linear projection of reaction center atom embeddings $C$; $K_j$, and $V_j$ are the linear projection of the concatenated molecule representations of reagents $H^{A}=[\mathbf{h}^{A}_1, \ldots, \mathbf{h}^{A}_{n_a}] \in \mathbb{R}^{n_a \times {d}}$. 
A cross attention layer has $n_h$ attention heads and $\text{HEAD}_j$ is the output from $j$-th attention head, $O \in \mathbb{R}^{d\times {d}}$ is a learnable parameter to integrate the attention heads.
The updated reaction center atom embeddings $\tilde{C}$ are obtained by passing $H$ to an FCN: 
\begin{equation}
\tilde{C}=\sigma\left(H W^c+B^c\right),
\end{equation}
where $W^c \in \mathbb{R}^{d\times {d}}$ and $B^c \in \mathbb{R}^{|C|\times d}$ are learnable parameters. After updating the reaction center atom embeddings in the reactants, we use \AI to derive the reactant molecule representations $\mathbf{h}_i^{R} \in \left\{ \mathbf{h}^{R}_1, \ldots, \mathbf{h}^{R}_{n_r}\right\}$.

\MIT uses a cross-attention layer to transform and integrate reagent information into the reaction center atoms, enabling the model to consider relationships between various reaction components. This makes \method learn a more chemically meaningful reaction representation by emphasizing reaction centers and reagent interactions.
We further show and analyze the benefits that \MIT brings to \method in Table~\ref{tbl:Ablation_study_module_2}, demonstrating its contribution to the overall performance of our model.

\subsection{Reaction Information Aggregation (\RIA)} 
After the derivation of representations for all the molecules involved in reactants, reagents, and products, we introduce \RIA to aggregate all the molecular information. This module explicitly describes the interaction of the involved molecules in the reaction and their contribution to yield.

Specifically, given the reactant molecule representations $\mathbf{h}_i^{R} \in \left\{ \mathbf{h}^{R}_1, \ldots, \mathbf{h}^{R}_{n_r}\right\}$, reagent molecule representations $\mathbf{h}_i^{A} \in \left\{ \mathbf{h}^{A}_1, \ldots, \mathbf{h}^{A}_{n_a}\right\}$, and the product molecule representations $\mathbf{h}_i^{P} \in \left\{ \mathbf{h}^{P}_1, \ldots, \mathbf{h}^{P}_{n_p}\right\}$, we first apply \MI to respectively derive three representations $\mathbf{r}$, $\mathbf{a}$, and $\mathbf{p}$ for reactant, reagent, and product. \MI uses a 
gating mechanism to aggregate the information from involved molecules. Taking reactant molecules as an example, this process can be described as follows:
\begin{equation}
\begin{aligned}
& \beta_i=\langle \mathbf{h}_i^{R} , \mathbf{w}_2 \rangle\\
& \mathbf{r}=\sum_{i=1}^{n_{r}} \mathbf{h}_i^{R} \times \beta_i,
\end{aligned}
\end{equation}
where $\mathbf{w}_2 \in R^{d}$ is a learnable vector to map each molecule representation to its weight $\beta_i$. 
This step allows \method to capture the collective properties within each group of molecules, providing a more compact and informative representation for subsequent processing.
These three representations are then concatenated to form a comprehensive representation of the entire reaction $\mathbf{x}=\left[\mathbf{r}, \mathbf{a}, \mathbf{p}\right] \in R^{3 d}$.  $\mathbf{x}$ then serves as the input for the yield predictor. 
%

\RIA processes reactant, reagent, and product molecules separately and aggregates information hierarchically to achieve a nuanced representation of the reaction. This design allows \method to capture each component's unique role and contribution to the reaction process, leading to a nuanced overall representation. 

\subsection{Yield Predictor} 
Provided with the comprehensive reaction representation $\mathbf{x}$, we stack two FCNs to predict the yield $\hat{y}$. The process is described below: 
\begin{equation}
\hat{y}=\left(\mathbf{x}\right)=f\left(\sigma\left(\mathbf{x} W_3+\mathbf{b}_1\right) \mathbf{w}_4+b_2\right),
\end{equation}
where $W_3 \in \mathbb{R}^{3d\times {d}}$, $\mathbf{w}_4 \in \mathbb{R}^{d}$, $\mathbf{b}_1 \in \mathbb{R}^{d}$ and $b_2\in \mathbb{R}^{1}$ are learnable parameters, and $f(\cdot)$ is a sigmoid function to control the predicted yield within the range $\left[0\%,100\%\right]$.


\subsection{Model training and hyperparameters optimization} 
During training, the mean absolute error (MAE) loss is optimized using adaptive moment estimation (Adam)~\cite{kingma2014adam}. 
\begin{equation}  
\label{Equation:mse}
\text { MAE }=\frac{\sum_{i=1}^N\left|y_i-\hat{y}_i\right|}{N}
\end{equation}
The initial learning rate is treated as a hyperparameter. 
We also utilize the validation set to schedule the learning rate decay patience and decay factor required in \textit{lr\_scheduler.ReduceLROnPlateau} provided by PyTorch~\cite{paszke2019pytorch}. 
All the searched hyperparameters and their respective search ranges are summarized in Table~\ref{tbl:hyperparameters_for_USPTO500MT} and Table~\ref{tbl:hyperparameters_for_BH}, respectively.
%

\section{Experiment results}

\subsection{Performance on the USPTO500MT dataset}

\subsubsection{Overall performance}
Table~\ref{tbl:models_on_USPTO500MT} presents the performance comparison of \methodwo, \methodw, and baseline methods \yieldbert and \chemt on the USPTO500MT dataset.
\input{tables/models_on_USPTO500MT}
\methodw demonstrates the best performance in terms of MAE and RMSE, achieving the lowest MAE of 0.179 and RMSE of 0.226. These results represent statistically significant improvements of 5.8\% on MAE over the previous best-performing method \chemt. The statistical significance of this improvement is underscored by a p-value of 5e-12
at a significance level of 5\%, obtained from a paired t-test comparing the Absolute Errors (AE) of \methodw and \chemt (Unless otherwise specified, the p-values mentioned in the following paper are all derived from this paired t-test).

\methodwo, which utilizes the basic atom features in contrast to \methodw utilizing the learned atom representations, achieved comparable results to \methodw with an MAE of 0.181. \methodwo is still significantly better than \chemt (p-value = 1e-8).
%
%
%
%
%
We attribute the superior performance to its effective framework design, specifically engineered to model and learn fundamental factors influencing reaction yield. The local-to-global learning scheme employed by \method allows for equal attention to all molecules of varying sizes before modeling their interactions, preventing the oversight of the contributions from small yet influential fragments (e.g., atoms, functional groups, or small molecules). This approach contrasts with sequence-based models like \chemt and \yieldbert, which treat the entire reaction as input, where the attention mechanisms may not be sufficiently sensitive to critical fragments.
Furthermore, \method's molecular interaction design explicitly models the function of reagents on reaction centers, more closely mimicking the synthetic reaction principle: reagents like catalysts have a huge impact on bond-breaking and formation. 
This targeted design is more effective than \chemt and \yieldbert's interaction modeling, which indiscriminately applies global attention to all atoms.
It is also worth noting that \methodwo is pre-training-free, whereas \chemt and \yieldbert are based on foundation models pre-trained on extensive molecule datasets (e.g. 97 million molecules from PubChem~\cite{kim2019pubchem}).
\methodwo's superior performance suggests that pre-training may not be necessary if the training dataset is sufficiently large (e.g., 78K for USPTO500MT) when the reactions are modeled in a targeted and explicit way. By incorporating more effective designs, \methodwo achieves better performance while saving huge resources required for pre-training.

\methodwo and \methodw exhibit nearly identical performance, with MAE values of 0.181 and 0.179, respectively. The former employs basic atom features, while the latter utilizes atom representations derived from the pre-trained MAT model \cite{maziarka2020molecule}.
The incorporation of learned representations does not obtain a substantial improvement in yield prediction accuracy over basic features. This outcome suggests that the atom representations acquired through the MAT model, which was originally developed for general molecule representation learning {\cite{maziarka2020molecule}}, lack the specificity required for reaction-oriented tasks.
%
%
%
Although basic atom features only provide elementary information about molecular properties, our findings underscore that the key to enhancing yield prediction accuracy lies in more sophisticated and effective modeling of intermolecular interactions.

While other graph-based yield prediction methods~\cite{saebi2023use,li2023reaction,yarish2023advancing} exist, they are primarily designed for datasets with fixed reagent structures, such as the Buchwald-Hartwig dataset, which includes very specific reagent information (additive, base, solvent, and ligand)~\cite{ahneman2018predicting}. However, these methods do not apply to the USPTO500MT dataset used in this study due to its varying number of reagents across reactions and lack of standardized reagent information.
%
%
However, the USPTO500MT dataset more closely resembles real-world scenarios where reaction compositions are not strictly structured. In this context, \method, \chemt, and \yieldbert demonstrate greater potential for practical applications compared to the graph-based methods just mentioned. 
\method's superior performance among those methods, as demonstrated in the previous results, combined with its flexibility in handling diverse inputs, positions it as a promising approach for accurate yield prediction in practical usage.
\subsubsection{Performance comparison over different yield ranges}
%
\input{tables/models_on_USPTO500MT_cate.tex}
\begin{figure}[h!]
\centering
\begin{subfigure}[t]{0.45\linewidth}
    \centering
    \includegraphics[width=\linewidth]{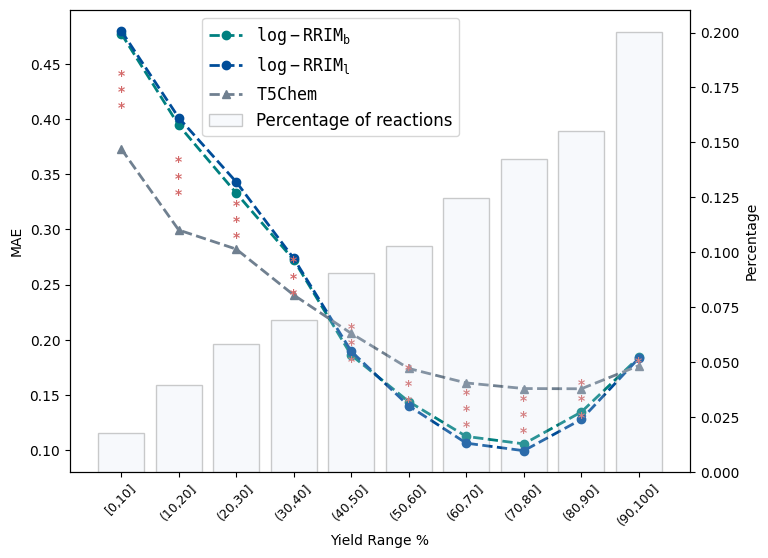}
    \caption[a]{Performance on reactions within each yield range}
    \label{fig:uspto_each_range}
\end{subfigure}\hfil
\begin{subfigure}[t]{0.45\linewidth}
    \centering
    \includegraphics[width=\linewidth]{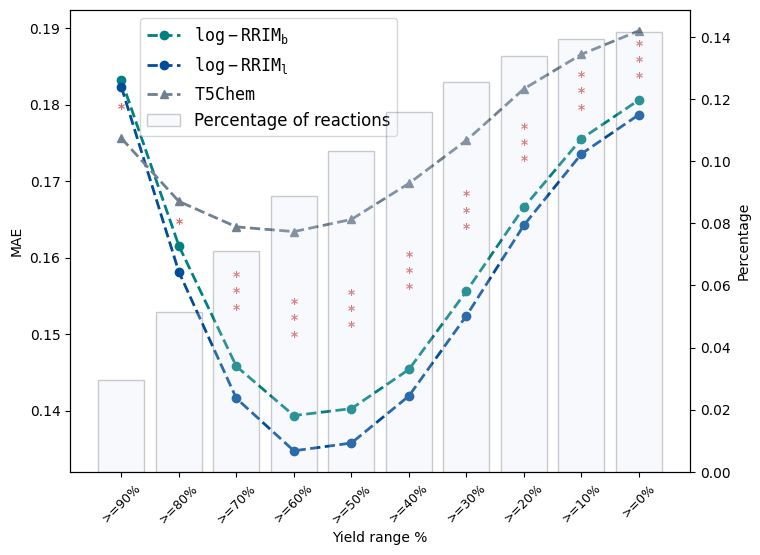}
    \caption[b]{Performance on reactions within cumulative yield range}
    \label{fig:uspto_accum_range}
 \end{subfigure}
\caption{Performance comparison of \method and \chemt across yield ranges on the USPTO500MT testing set. Left y-axis: MAE of predicted yields. Right y-axis: percentage of reactions in the testing set for each yield range. 5\% significance level: * for $\text{p-values}<0.05$, ** for $\text{p-values}<0.005$, *** for $\text{p-values}<0.0005$.}
\label{fig:USPTO_range}
\end{figure}

To gain deeper insights into the performance differences between \method and \chemt, we conducted a detailed analysis of predictions across various yield ranges.
Following Yarish \etal~\cite{yarish2023advancing}'s reaction yield categorization, we summarize the MAE of \method compared to \chemt across low, medium, and high-yielding reactions in Table~\ref{tbl:models_on_USPTO500MT_cate}. As shown in the table, both \methodwo and \methodw notably outperform \chemt on medium-to-high yielding reactions, with MAE improvements up to 12.6\%.
Figure~\ref{fig:USPTO_range} provides a more detailed visualization of these comparisons by dividing yields into 10\% intervals, with stacked asterisks indicating the statistical significance of performance differences across yield ranges (see Table~\ref{tbl:USPTO_model_comparison} for exact values). 
Figure~\ref{fig:uspto_each_range} shows that both \methodwo and \methodw outperform \chemt in predicting yields within the 40\% to 100\% with t-test p-values all less than 0.05, indicating statistical significance at the 5\% level. 
This pattern suggests that \method is a more reliable predictor for medium to high-yielding reactions, a crucial advantage in practical synthesis scenarios~\cite{ma2024we,kawasaki1999high}.
Also, Figure~\ref{fig:uspto_accum_range} suggests the overall prediction performance of \method is significantly better. This improved overall accuracy is particularly valuable in the context of exploring new reactions, where precise yield data may not be available for reference. In such scenarios, \method's overall more reliable predictions can offer more accurate guidance for reaction planning and optimization. 
However, for reaction yields below 40\%, \method exhibits inferior performance than \chemt. We attributed this to \chemt's leveraging of foundation models pre-trained on extensive molecule datasets, compensating for a potential shortage of training samples 
encountered by \method on reactions with yields below 40\%
(18.1\% of the training set). 
Nevertheless, \method remains the preferred choice for chemists seeking reliable yield predictions, particularly for medium to high-yielding reactions or when no preliminary reaction yield data can be referred to. 
This reliability can significantly aid chemists in experimental planning, reducing the number of optimization iterations and minimizing resource consumption. 
\subsubsection{Effectiveness in reactant-reagent interactions modeling}
%
To assess the model's capacity to capture the influence of molecular interactions on yield, specifically how reactants and reagents affect each other in the context of a reaction, we conducted two analyses on the testing set of USPTO500MT.
First, we identified 76 reaction pairs (152 reactions) with identical reactants but different reagents and yields. This setup allowed us to evaluate how our method is sensitive to the effects of reagents on yields.
In this context, "interactions" refer to how the introduction of different reagents influences the reaction outcome with the same reactants.
\methodwo achieved a prediction MAE of 0.145, notably outperforming \chemt's 0.182 among those reactions. Furthermore, \methodwo correctly predicted the yield difference (how much the yield increases or decreases) in 62\% (47 out of 76) of reaction pairs, compared to \chemt's 38\%. 
This suggests that \methodwo is more sensitive to reagent changes and their effects on yield.
Case 1 in Figure~\ref{uspto_case_study} illustrates this: in two identical aryl nitration reactions, adding ether as a solvent increases the ground-truth yield from 42.0\% to 57.7\%. \methodwo correctly predicts this upward trend, while \chemt does not. This shows \methodwo's ability to capture how the addition of a solvent (ether) interacts with the existing reactants to influence the yield.

Secondly, we examined 3,698 reactions grouped into 619 sets, each containing two or more reactions with identical reagents but different reactants. 
This analysis aimed to evaluate the models' ability to predict yields when the same reagents interact with various reactants. Here, "interactions" refer to how the same set of reagents behaves differently with varying reactants.
%
%
\methodwo exhibited more accurate predictions in 58\% of sets (357 out of 619), with a lower MAE of 0.147 compared to \chemt's 0.222.
Case 2 in Figure~{\ref{uspto_case_study}} demonstrates \methodwo's consistently more accurate predictions when the same reagents (carbon disulfide and bromine) interact with two different reactants. 
This indicates \methodwo's enhanced capability to learn and model specific reagent functions across different reaction contexts, capturing how the same reagents behave differently with varying reactants.

Overall, These analyses suggest that \methodwo is more sensitive to changes in reactant-reagent combinations, indicating better modeling of their interactions. 
This enhanced capability makes \methodwo a potential aid for chemists in selecting and optimizing reactants or reagents during synthesis planning.
We attribute this superiority to \methodwo's explicit modeling of reagent function to the reaction center. This approach, implemented through a cross-attention mechanism, 
aligns with fundamental reaction principles. It allows \methodwo to directly model how reagents influence the reaction center, providing a more nuanced understanding of the reaction process. 
An ablation study on the removal of explicit reagent function modeling, provided in Table~\ref{tbl:Ablation_study_module_2}, further supports this design choice. 
As a result, \methodwo demonstrates an enhanced ability to capture and interpret complex reactant-reagent interactions, leading to more accurate yield predictions across diverse reaction component combinations.

%
\subsubsection{Sensitivity to small fragments modifications} 
To evaluate the models' ability to capture the influence of involved small fragments on reaction yields, we conducted a comparative analysis of their performance on similar reactions with small differences only on a few small fragments in reactants or reagents. 
Given the absence of a standardized method for quantifying reaction similarity, we propose a novel similarity metric $ Sim(\mathcal{X}_i, \mathcal{X}_j)$ between reactions $\mathcal{X}_i$ and $\mathcal{X}_j$, defined as the average of reactant and reagent similarities:
%
\begin{equation}
\label{Equation:r2}
Sim(\mathcal{X}_i, \mathcal{X}_j) =\frac{1}{2}\left[s\left({\mathcal{R}}_i, {\mathcal{R}}_j\right)+s\left({\mathcal{A}}_i, {\mathcal{A}}_j\right)\right]
\end{equation}
where $\mathcal{X}: \mathcal{R} \xrightarrow[]{\mathcal{A}} {\mathcal{P}}$ refers to the reaction, $\mathcal{R}$ and $\mathcal{A}$ are the concatenation of all reactants and reagents in the reaction, respectively. $s(\cdot, \cdot)$ is the Tanimoto coefficient between the two chemical structures of Morgan fingerprint~\cite{morgan1965generation}. 
%

\begin{figure*}[h!]
\begin{center}
{\includegraphics[scale=0.6]{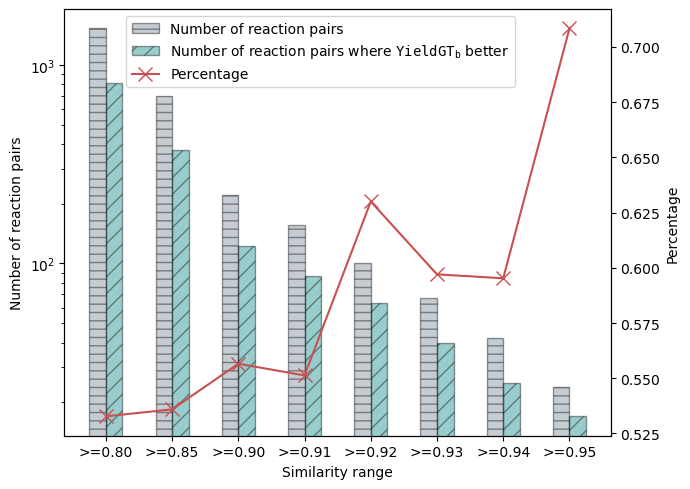}}
\caption{Model performance on reaction pairs categorized by similarity. The left y-axis displays the number of reaction pairs on a logarithmic scale. Grey bars indicate the number of reaction pairs within each similarity range. Green bars represent the number of reaction pairs where \methodwo predicts more accurately than \chemt. The right y-axis shows the percentage of reaction pairs with more accurate predictions by \methodwo relative to the total number of reactions in each similarity range, as depicted by the red line. }
\label{Similarity}
\end{center}
\end{figure*}

We evaluated reaction pairs across a range of similarity thresholds (0.8-0.95), comparing the performance of \methodwo and \chemt in predicting yield differences between the two reactions in the pair. The results are illustrated in Figure~\ref{Similarity}.
Specifically, for reaction pairs with $Sim\geq 0.80$ (1526 pairs, 3052 reactions), {\methodwo} outperformed {\chemt} on 53\% (813/1526) pairs, with overall MAEs of 0.158 and 0.159 respectively.
%
%
This advantage becomes more pronounced as the reaction similarity increases. On 221 pairs with $Sim\geq 0.9$, \methodwo surpassed \chemt on 56\% (123/221), with MAEs of 0.162 and 0.165 respectively. The trend culminated with highly similar reaction pairs ($Sim\geq 0.95$, 24 pairs), where \methodwo demonstrated marked superiority, outperforming \chemt on 71\% (17/24), with MAEs of 0.150 and 0.170 respectively.
These results reveal a clear trend: \methodwo's accuracy in capturing yield differences improves as reaction similarity increases.
This indicates that \methodwo exhibits enhanced sensitivity to subtle component changes that impact reaction yields, particularly for highly similar reactions. 

\begin{figure}[h!]
\begin{center}
{\includegraphics[scale=0.55]{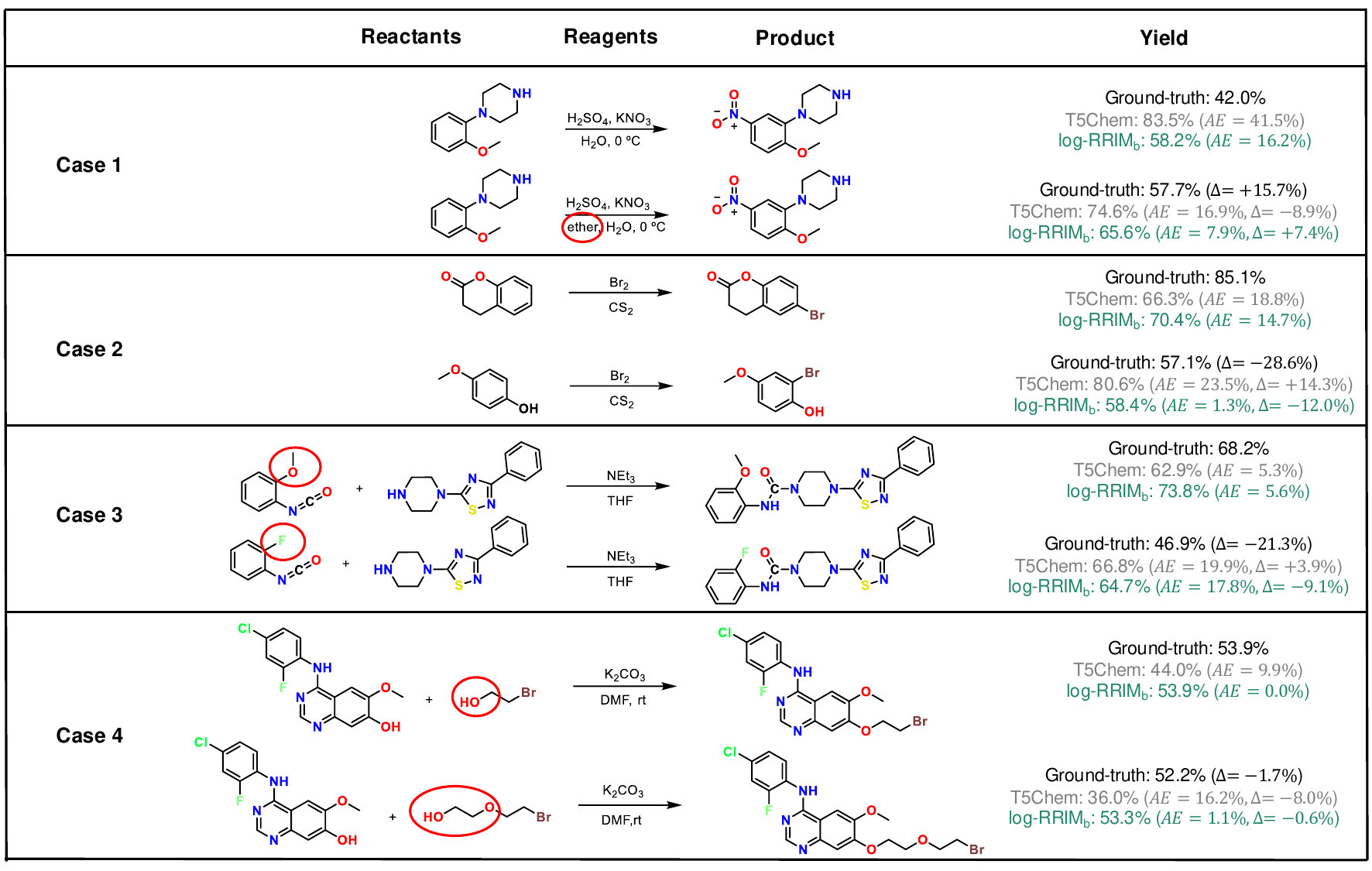}}
\caption{Cases analysis on the USPTO500MT dataset. Each reaction is reported with reactants, reagents, products, and the ground-truth and predicted yields by \chemt and \methodwo. $AE$ in parentheses represents the \underline{a}bsolute \underline{e}rror between the predicted and ground-truth yields. $\Delta$ in parentheses represents the change of the ground-truth and predicted yields in the second reaction to the corresponding value in the first reaction.
%
}
\label{uspto_case_study}
\end{center}
\end{figure}

The capability is also demonstrated in several cases. In Figure~\ref{uspto_case_study} case 3, the two reactions differ only in their \textit{ortho}-substitution (methoxy vs fluoro group), resulting in a yield decrease from 68.2\% to 48.9\%. \method correctly predicts this change, while \chemt incorrectly predicts the opposite trend. 
Similarly, case 4 in Figure~\ref{uspto_case_study} presents two alkylations of hydroxyquinoline with different alkylating agents. The ground-truth yield changes minimally in this situation, which \methodwo correctly predicts, whereas \chemt makes an erroneous prediction.
These results indicate that the \methodwo excels in predicting yield changes triggered by those small modifications in atoms, functional groups, or small molecules in reactants or reagents. 
This capability is essential for optimizing reactions in complex chemical systems, where small adjustments to reactants and reagents can significantly impact yields. \methodwo's precision in predicting the effects of these subtle changes enhances its utility for guiding synthetic strategies and fine-tuning reactions. By offering reliable forecasts for small modifications, \methodwo can potentially streamline the optimization process, reducing the number of experimental iterations required and saving time and resources in research and industrial settings.

This capability stems from \methodwo's unique local-to-global learning strategy. By first analyzing each molecule separately and then modeling their interactions, the model ensures equal consideration of all molecules, regardless of their size. 
This approach differs from sequence-based models like \chemt, which process the entire reaction SMILES string simultaneously. Such models may overlook crucial smaller fragments that significantly impact the overall yield, as the global attention mechanisms might not be sufficiently sensitive to these critical molecular fragments.

Overall, the performance differences presented in these analyses underscore our belief that model frameworks should be carefully designed based on specific task characteristics rather than solely relying on foundation models.
While they have shown great promise in many areas, a basic fine-tuning strategy may not always be optimal for specialized tasks like reaction yield prediction. Such an approach lacks task-specific module designs that capture the intricate characteristics of chemical reactions, potentially limiting the performance. 

\subsection{Performance on the external dataset CJHIF}
To assess our model's performance on external
datasets, we conducted an evaluation using a subset of the CJHIF dataset~\cite{jiang2021smiles}.
This approach involves using models trained on USPTO500MT and testing them on a subset of the CJHIF dataset, which comprises 3,219,165 reactions sourced from high-impact factor journals.
Our assessment involved 1,000 non-zero-yielding chemical reactions randomly selected from the initial 50,000 reactions in the CJHIF dataset.
We specifically chose reactions with reported non-zero yields because CJHIF treats unreported yields as zeros, and we aimed to evaluate our model on reactions with confirmed, measurable outcomes. 
Importantly, these 1,000 reactions are not included in the training or testing data of USPTO500MT, thus providing an independent testing set for assessing our model's performance on external reactions.
%

\begin{figure}[h!]
\begin{center}
{\includegraphics[scale=0.6]{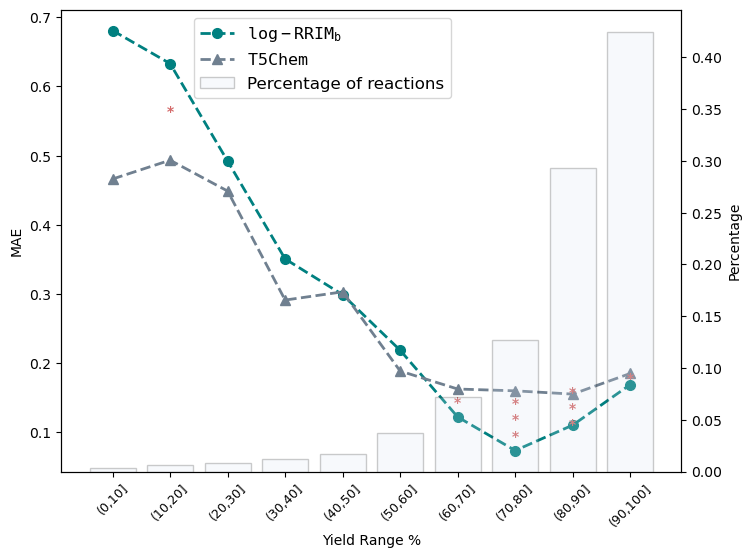}}
\caption{\methodwo and \chemt performance comparison over each yield range on a subset of CJHIF. Left y-axis: MAE of predicted yields. Right y-axis: percentage of reactions in the testing set for each yield range. 5\% significance level: * for $\text{p-values}<0.05$, ** for $\text{p-values}<0.005$, *** for $\text{p-values}<0.0005$.}
\label{fig:YieldGT_and_T5Chem_CJHIF}
\end{center}
\end{figure}

%
Overall, \methodwo achieved an MAE of 0.149, representing a 16.8\% improvement over \chemt's MAE of 0.179. The results of analyzing performance across yield ranges are illustrated in Figure~\ref{fig:YieldGT_and_T5Chem_CJHIF}. 
\methodwo significantly outperformed \chemt for reactions with yields between 60\% to 100\% (confidence level 95\%, more details are provided in Table~{\ref{tbl:t_test_yield_ranges_CJHIF}}). 
This superior performance aligns closely with our observations from the USPTO500MT dataset, particularly in \methodwo's enhanced accuracy for medium to high-yielding reactions, which suggests that \methodwo's improved predictive power for high-yielding reactions is a generalizable feature, not limited to a specific dataset.
We attribute this generalizability to \methodwo's molecular interaction design which uses the cross-attention mechanism to effectively model the function of reagents in relation to the reaction center. This allows \methodwo to learn fundamental principles about how reagents impact bond-breaking and formation, which are key factors affecting reaction yield. 
The extensive data in USPTO500MT training data enables \methodwo to learn such principles to achieve better test performance on external datasets.

\begin{figure}[h!]
\begin{center}
\includegraphics[scale=0.5]{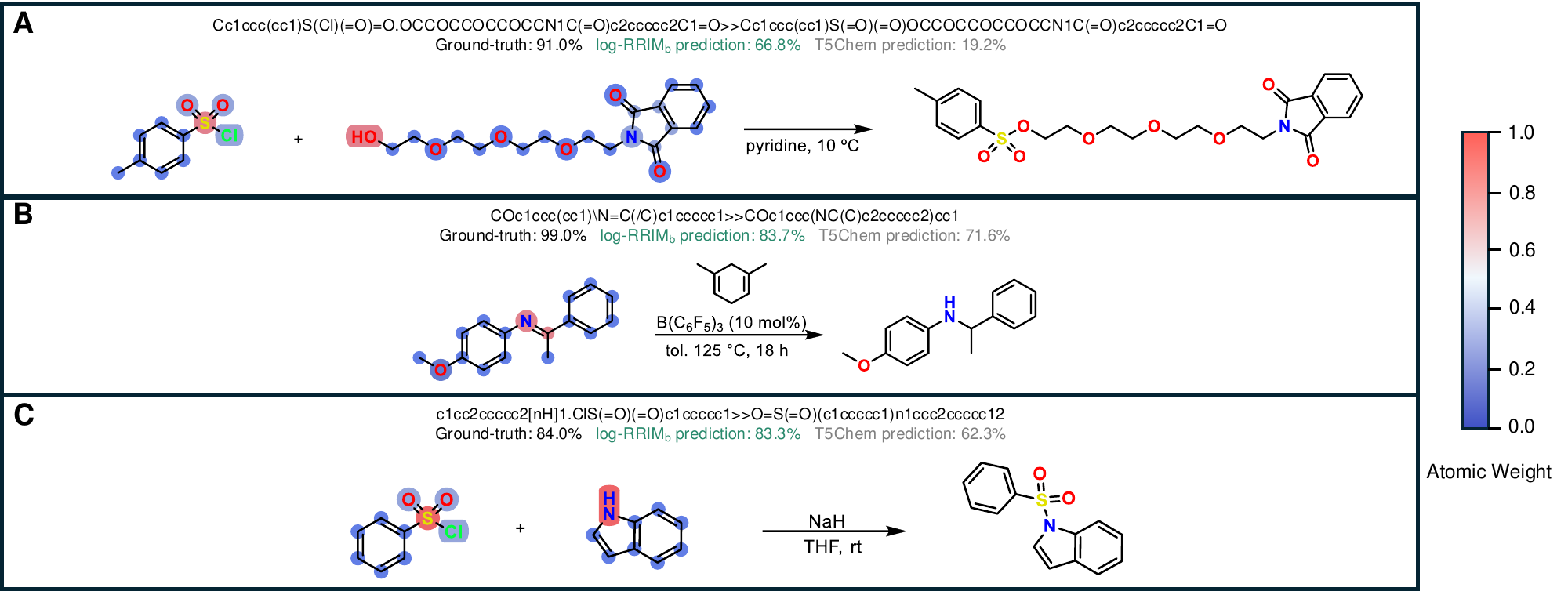}
\caption{Visualizations of atom contribution (atomic weight) in learning molecule representation. The contribution is quantified by the color and the three exemplar reactions are selected from the CJHIF dataset.}
\label{fig: atom_weight_visualization}
\end{center}
\end{figure}

To further validate that \method has effectively learned key factors influencing reaction yield, we visualized the contribution (weight) of each atom when \methodwo aggregates atom embeddings and constructs the molecule representation. Three exemplar reactions are shown in Figure~\ref{fig: atom_weight_visualization}.
In reaction \textbf{A}, a sulfonylation reaction, the sulfur-bearing sulfonyl chloride group on the \textit{p}-toluenesulfonyl chloride and the free hydroxyl (OH) group on the alcohol are the two reacting centers. The oxygen (O) acts as the nucleophile that displaces the chlorine (Cl) atom, and these atoms influence the yield of the reaction.
Reaction \textbf{B} is an imine reduction reaction of the compound N-(4-methoxyphenyl)-1-phenylethylamine. The polar C=N bond between the Nitrogen (N) and Carbon (C) is the reactive site, and these two atoms influence the yield of the reaction, which results in the single C-N bond in the corresponding amine.
In reaction \textbf{C}, the two atoms that ultimately influence the yield are Sulfur (S) of benzene sulfonyl chloride and Nitrogen (N) of the indole, producing the final compound.
Combined with the weights highlighted by the colormap in Figure~\ref{fig: atom_weight_visualization}, we found that the atoms mentioned above that have a greater impact on yield are given higher weights by \methodwo, and these atoms are also the atoms in the reaction center.
This finding aligns with the fundamental chemical principle that reaction center atoms play a crucial role in the bond-breaking and bond-forming steps in the transition state, thereby exerting substantial influence on the yield. 
The ability of \methodwo to prioritize these critical atoms in learning the molecule representation is essential for building more accurate models for predicting reaction yield, and our method demonstrates particular effectiveness in this regard.
\subsection{Performance on the Buchwald-Hartwig dataset}
On the Buchwald-Hartwig dataset, we conducted a performance comparison among pre-training-free models (\yieldgnn, \semg, and \rdmpnn), using 10-fold cross-validation~\cite{geisser1975predictive}, and reported the testing results averaged over the 10 folds.
Table~\ref{tbl:Pretraining-free_models_on_BH} reports the mean and standard deviation (in parentheses) of MAE, RMSE, and $R^2$.
Table~\ref{tbl:model_on_each_data_split} provides a detailed comparison of testing MAE values for different models on each fold.
Our method, \methodwo, outperforms other pre-training-free (also graph-based) methods across all evaluation metrics (MAE 0.0348, RMSE 0.0544, and $R^2$ 0.953). Notably, it achieves a 14.7\% improvement in MAE over the best-performing baseline, \yieldgnn.
We attribute \methodwo's superior performance to its more effective molecular interaction design, explicitly modeling the reagents' function to the reaction center. 
By incorporating this design, \methodwo captures crucial chemical insights that other methods may overlook, leading to more accurate predictions.
%
Compared to the second-best baseline method, \semg, \methodwo improves the MAE by 17.9\%. To put this improvement in context, it is worth recalling that \semg focuses on building more informative atom features (digitalized steric and electronic information). In contrast, \methodwo emphasizes learning the characteristics of the reaction itself and molecular interactions. The performance difference between these approaches suggests that for yield prediction tasks, the latter strategy may be more effective.
In summary, \methodwo outperforms other pre-training-free and graph-based models substantially.
These results demonstrate the importance of focusing on reaction characteristics and molecular interactions in yield prediction tasks.

As shown in Table~\ref{tbl:Pretraining_models_on_BH}, when compared to the pre-training-based methods, which are also sequence-based models (\chemt and \yieldbert), \methodw also shows competitive performance by achieving the MAE of 0.0347, which has a 16.4\% improvement over \yieldbert but inferior to \chemt by 11.6\%. These results indicate our method is comparable to the best-performing sequence-based model \chemt while using only 2\% of the pre-training dataset size compared to \chemt (2M vs 97M).
To further validate this comparison, we conducted statistical analysis on the predicted values of \methodw and \chemt for each yield range on the first data split (with detailed p-values shown in Table~\ref{tbl:YieldGT_T5Chem_comparison_BH}).
The analysis shows that, for all ranges except the 10\%-20\%, 40\%-50\%, and 70\%-80\% ranges (a total of 27.6\% of the test reactions), the differences between \methodw and \chemt are not statistically significant at the 95\% confidence interval. This indicates that the performance of \methodw and \chemt is largely comparable across most yield ranges.
Note that the Buchwald-Hartwig dataset involves only a single reaction type with limited components. 
On this specific reaction type, \method could underperform \chemt. 
However, real-world chemical synthesis generally involves reactions of multiple types. Thus, methods that could accurately predict yields of various types are highly demanded.
As shown on the USPTO500MT and CJHIF datasets, which contain numerous reaction types, our method demonstrates superior performance. On these more diverse and complex datasets, \method outperforms \chemt. 
These results demonstrate the potential superior utility of \method over \chemt in real-world chemical synthesis applications.

We also note that the performance of \methodwo and \methodw are nearly identical (MAE 0.0348 v.s 0.0347) on the Buchwald-Hartwig dataset. 
This observation aligns with the results we obtained on the USPTO500MT dataset. 
These consistent findings across different datasets suggest that effective molecular interaction modeling may play a more crucial role than using pre-trained models to generate informative atom representations in yield prediction tasks.
%
%
%
\input{tables/Pretraining-free_models_on_BH}
\input{tables/Pretraining_models_on_BH}

\begin{figure}[h!]
\begin{center}
{\includegraphics[scale=0.6]{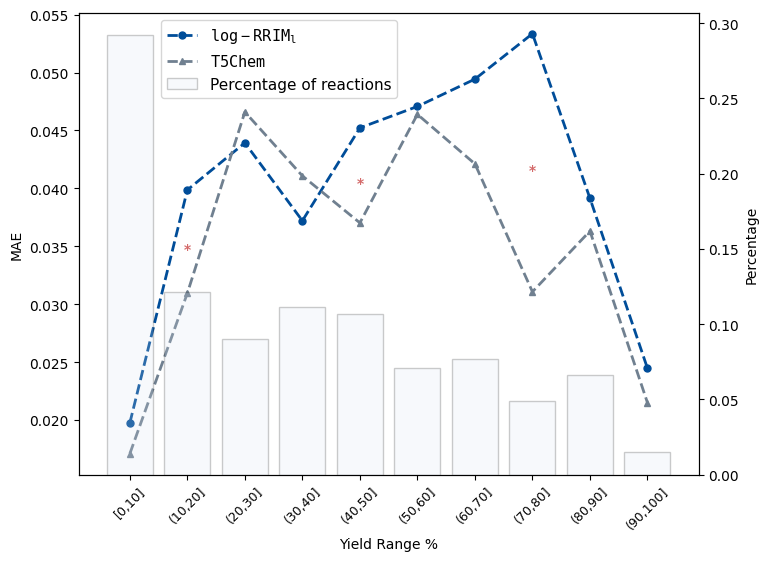}}
\caption{Performance comparison of \methodw and \chemt across yield ranges on the first data split of the Buchwald-Hartwig dataset. Left y-axis: MAE of predicted yields. Right y-axis: percentage of reactions in the testing set for each yield range. 5\% significance level: * for $\text{p-values}<0.05$, ** for $\text{p-values}<0.005$, *** for $\text{p-values}<0.0005$.}
\label{fig:bh_each_range}
\end{center}
\end{figure}
\section{Discussion} 
%

In conclusion, in this paper, we present \method, a novel graph-transformer-based reaction representation learning framework for yield prediction. \method leverages a local-to-global representation learning process and incorporates a cross-attention mechanism to model reagent-reaction center interactions, facilitating improved capture of small fragment contributions and interactions between reactant and reagent molecules. This approach allows \method to tap into crucial aspects of chemical knowledge, particularly the importance of reagent effects and reaction center dynamics in determining reaction outcomes.
Without reliance on pre-training tasks, \method demonstrates superior accuracy and effectiveness compared to other graph-based methods and state-of-the-art sequence-based approaches, particularly for medium to high-yielding reactions. Our analyses further show \method's advanced modeling of reactant-reagent interactions and sensitivity to small molecular fragments, making it a valuable asset for reaction planning and optimization in chemical synthesis.

The \method framework requires that predicted reactions consist of three parts (reactant, reagent, and product) and that reaction center atoms be correctly identifiable. While this may limit its practical applications in some scenarios, it enables \method to more effectively model the crucial intermolecular dynamics that significantly influence reaction outcomes. This approach underscores the importance of incorporating chemical-specific information into model architecture design, rather than directly adapting general-purpose foundation models for chemical tasks like yield prediction.

While \method makes significant strides in leveraging chemical knowledge, particularly in modeling reagent-reaction center interactions, there remains a vast body of chemical expertise that could potentially be incorporated to further enhance the performance. 
For instance, research has elucidated detailed mechanisms for different reaction types, like transition states~\cite{carey2007advanced, anslyn2006modern}, which are not yet explicitly incorporated into our model. Furthermore, chemists have developed a deep understanding of the relative reactivity of different functional groups under various conditions~\cite{clayden2012organic, smith2020march}, which represents another rich source of knowledge that could be integrated into the model.
Incorporating such additional aspects of chemical knowledge presents both a challenge and an opportunity for future research. It could potentially enhance the model's predictive power, improve its generalization to diverse reaction types, and provide more interpretable insights into the factors driving yield predictions. 
Another promising direction for future research is the exploration of multi-task learning approaches, where the model could be trained simultaneously on yield prediction, reaction condition optimization, retrosynthesis planning, etc. This could lead to a more comprehensive understanding of chemical reactivity and potentially improve performance across all tasks.

\method represents a significant step forward in reaction yield prediction by leveraging graph-based representations and modeling reagent-reaction center interactions, and there is still room for further integration of chemical knowledge and enhancement of the model’s capabilities. By continuing to merge data-driven techniques with established chemical principles, it is crucial to develop more robust, versatile, and reliable models for computational chemistry.

\section{Data and Software Availability}
The data used and the code for \method are made publicly available at~\href{https://github.com/ninglab/Yield_log_RRIM}{\nolinkurl{https://github.com/ninglab/Yield\_log\_RRIM}}.

\section{Acknowledgements}
This project was made possible, in part, by support from the
National Science Foundation grant no. IIS-2133650 (X.N.)
and the National Library of Medicine grant no.
1R01LM014385-01 (X.N., D.A.). Any opinions, findings and
conclusions or recommendations expressed in this paper are
those of the authors and do not necessarily reflect the views of
the funding agency.

\clearpage

\bibliography{refs}

\clearpage
\section{Supporting Information}
\subsection{Exact paired t-test p-values between \method and \chemt on each dataset}
The tables below include all the MAE differences and p-values of the paired t-tests between \methodwo and \chemt.
\input{tables/t_test_yield_ranges}
\input{tables/t_test_yield_ranges_CJHIF}
\input{tables/t_test_yield_ranges_BH}

\subsection{Performance on each Buchwald-Hartwig data split}
In Table~\ref{tbl:model_on_each_data_split}, we present the performance (MAE on the testing set) of each model across every data split.
\input{tables/model_on_each_data_split}

\subsection{Impact of explicit reagents function modeling}
We conducted an ablation study to investigate the importance of explicitly modeling reagent effects on reaction yield by removing the second module Molecular Interaction (\MIT) from our proposed \methodwo framework. 
In this modified version, atom embeddings of the reactant molecules are directly passed to the Atomic Interaction (\AI) module to derive reactant molecule representations without using the cross-attention mechanism to update the reaction center atom embeddings.
We compared the performance of this ablated model with the original \methodwo on the first data split of the Buchwald-Hartwig dataset. The results are summarized in Table~\ref{tbl:Ablation_study_module_2}.
Results show that removing the \MIT module hugely decreased prediction performance, with the MAE of \methodwo increasing by 45.0\%. This substantial drop in accuracy underscores the critical role of explicitly modeling reagent effects in yield prediction tasks.
The observed performance degradation can be attributed to the \MIT module's ability to capture fundamental characteristics of chemical reactions, particularly the influence of substances such as catalysts on bond breaking and formation at reaction centers. 
By incorporating this knowledge, \method enhances its capacity to construct more informative molecular and reaction representations, ultimately making more accurate reaction yield predictions.
\input{tables/Ablation_study_module_2}

\subsection{Hyperparameters}
Table~\ref{tbl:hyperparameters_for_USPTO500MT} and 
Table~\ref{tbl:hyperparameters_for_BH} summarizes the searched hyperparameters and their ranges on the USPTO500MT and Buchwald-Hartwig datasets. The selected values for \methodwo are highlighted by underlining, while those for \methodw are indicated in bold. Table~\ref{tbl:pretrained_MAT_hyperparameters} summarizes the hyperparameters used in the pre-trained MAT model.
\input{tables/hyperparameters_for_USPTO500MT}
\input{tables/hyperparameters_for_BH}
\input{tables/pretrained_MAT_hyperparameters}

\subsection{SMILES}
In this section, we provide all the SMILES of compounds we mentioned in the manuscripts in Table~\ref{tbl:SMILES}.
\input{tables/SMILES.tex}

\clearpage

\begin{figure}[h!]
  \centering
  \includegraphics[height=1.75in]{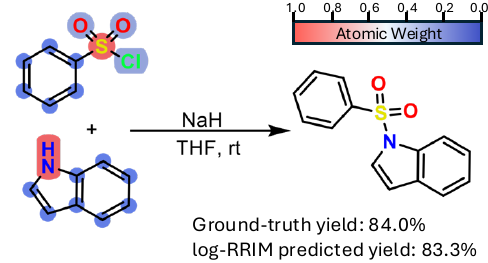}
  \\
  \vspace{10pt}
  For Table of Contents Only
\end{figure}

\end{document}

%% file: define.tex
\newcommand{\etal}{\mbox{\emph{et al.}}\xspace}

\newcommand{\method}{\mbox{$\mathop{\mathsf{log\text{-}RRIM}}\limits$}\xspace}
\newcommand{\methodwo}{\mbox{$\mathop{\mathsf{log\text{-}RRIM_{b}}}\limits$}\xspace}
\newcommand{\methodw}{\mbox{$\mathop{\mathsf{log\text{-}RRIM_{l}}}\limits$}\xspace}
\newcommand{\MRL}{\mbox{$\mathop{\mathsf{MRL}}\limits$}\xspace}
\newcommand{\AI}{\mbox{$\mathop{\mathsf{AI}}\limits$}\xspace}
\newcommand{\MI}{\mbox{$\mathop{\mathsf{MI}}\limits$}\xspace}
\newcommand{\MIT}{\mbox{$\mathop{\mathsf{MIT}}\limits$}\xspace}
\newcommand{\RIA}{\mbox{$\mathop{\mathsf{RIA}}\limits$}\xspace}

\newcommand{\chemt}{\mbox{$\mathop{\mathsf{T5Chem}}\limits$}\xspace}
\newcommand{\yieldbert}{\mbox{$\mathop{\mathsf{YieldBERT}}\limits$}\xspace}
\newcommand{\yieldgnn}{\mbox{$\mathop{\mathsf{YieldGNN}}\limits$}\xspace}
\newcommand{\rdmpnn}{\mbox{$\mathop{\mathsf{RD\text{-}MPNN}}\limits$}\xspace}
\newcommand{\semg}{\mbox{$\mathop{\mathsf{SEMG\text{-}MIGNN}}\limits$}\xspace}

%% file: tables/Atom_features.tex
\begin{table}[htbp]
\caption{Basic atom features used in \method}
\label{tbl:Atom_features}
\centering
\begin{threeparttable}
\begin{tabular}{
    @{\hspace{8pt}}l@{\hspace{8pt}}
    @{\hspace{2pt}}l@{\hspace{2pt}}
    }
    \toprule
    \textbf{Indices} & \textbf{Description} \\
    \midrule
    
    0-11 & Atom type of B, N, C, O, F, P, S, CL, BR, I, Dummy, Other (One-hot encoded) \\
    12-17 & Number of connected heavy atoms of 0, 1, 2, 3, 4, 5 (One-hot encoded)\\
    18-22 & Number of connected hydrogen of 0, 1, 2, 3, 4 (One-hot encoded)\\
    23-25 & Formal charge of -1, 0, 1 (One-hot encoded) \\
    26 & If the atom is in a ring (Binary)\\
    27 & If it is aromatic (Binary)\\
    \bottomrule
    \end{tabular}
\end{threeparttable}
\end{table}

%% file: tables/notations.tex
\begin{table}[htbp]
\caption{Key Notations}
\label{tbl:Notations}
\centering

\begin{threeparttable}
\begin{tabular}{
    @{\hspace{8pt}}l@{\hspace{10pt}}
    @{\hspace{2pt}}l@{\hspace{2pt}}
    }
    \toprule
    \textbf{Notation} & \textbf{Meaning} \\
    
    \midrule
    
    $\mathcal{X}$ & Reaction \\
    $\mathbf{x}$ & Reaction representation\\
    $y$, $\hat{y}$ & Ground truth reaction yield and the predicted yield\\
    
    \midrule
    
    $\mathcal{R}$, $\mathcal{A}$, $\mathcal{P}$ & Reactant, reagent and product\\
    $\mathbf{r}, \mathbf{a}, \mathbf{p}$ & Reactant, reagent, and product representation\\ 
    $n_r, n_a, n_p$ & Number of molecules in the reactant, reagent, and product\\
    
    \midrule
    
    $\mathcal{M}$ & Molecule\\
    $n$ & Number of atoms of the molecule\\
    $M$ & Atom embeddings of the molecule\\
    $\mathbf{h}_m$ & Molecule representation\\

    \bottomrule
    \end{tabular}
\end{threeparttable}
\end{table}

%% file: tables/models_on_USPTO500MT.tex
\begin{table}[htbp]
\caption{Model performance comparison on USPTO500MT}
\label{tbl:models_on_USPTO500MT}
\centering
\begin{threeparttable}
\begin{tabular}{    
    @{\hspace{10pt}}c@{\hspace{10pt}}
    @{\hspace{10pt}}c@{\hspace{10pt}}
    @{\hspace{10pt}}c@{\hspace{10pt}}
    @{\hspace{10pt}}c@{\hspace{10pt}}
    }
\toprule
\textbf{Method} & \textbf{MAE} & \textbf{RMSE} & $\mathbf{R^2}$ \\
\midrule
\yieldbert & 0.191 & 0.245 & 0.090\\
\chemt & 0.190 & 0.249 & \textbf{0.212} \\
\methodwo & \textbf{0.181} & \textbf{0.228} & 0.122  \\
\methodw & \textbf{0.179} & \textbf{0.226} & 0.144 \\
\bottomrule
\end{tabular}

\begin{tablenotes}
\begin{footnotesize}
\item 
Ground-truth yield ranging from 0-1 (0\%-100\%). The best performance in each range is highlighted in bold.
      \par
\end{footnotesize}
\end{tablenotes}

\end{threeparttable}

\end{table}

%% file: tables/models_on_USPTO500MT_cate.tex
\begin{table}[htbp]
\caption{Model performance comparison on USPTO500MT with different yield range}
\label{tbl:models_on_USPTO500MT_cate}
\centering
\begin{threeparttable}
\begin{tabular}{    
    @{\hspace{10pt}}c@{\hspace{10pt}}
    @{\hspace{10pt}}c@{\hspace{10pt}}
    @{\hspace{10pt}}c@{\hspace{10pt}}
    @{\hspace{10pt}}c@{\hspace{10pt}}
    @{\hspace{10pt}}c@{\hspace{10pt}}
    }
\toprule
\multirow{2}{*}{\textbf{Method}}
\textbf{}&\multicolumn{3}{c}{\textbf{Yield Ranges}} \\
\cmidrule{2-4} 
 & \textbf{Low: [0\%-33\%]} & \textbf{Medium: (33\%-66\%]} & \textbf{High: (66\%-100\%]} \\
\midrule
\chemt & \textbf{0.294} & 0.191 & 0.161 \\
\methodwo & 0.364 & 0.169 & 0.142  \\
\methodw & 0.373 & \textbf{0.167} & \textbf{0.138} \\
\bottomrule
\end{tabular}

\begin{tablenotes}
\begin{footnotesize}
\item 
The values are MAE of predictions on reactions whose ground-truth yields fall within each yield range. The best performance in each range is highlighted in bold.
      \par
\end{footnotesize}
\end{tablenotes}

\end{threeparttable}

\end{table}

%% file: tables/Pretraining-free_models_on_BH.tex
\begin{table}[htbp]
\caption{Pre-training-free models performance comparison on the Buchwald-Hartwig dataset}
\label{tbl:Pretraining-free_models_on_BH}
\centering
\begin{threeparttable}
\begin{tabular}{    
    @{\hspace{10pt}}c@{\hspace{10pt}}
    @{\hspace{10pt}}c@{\hspace{10pt}}
    @{\hspace{10pt}}c@{\hspace{10pt}}
    @{\hspace{10pt}}c@{\hspace{10pt}}
    }
\toprule
\multirow{2}{*}{\textbf{Method}}
\textbf{}&\multicolumn{3}{c}{\textbf{Metrics}} \\
\cmidrule{2-4} 
\textbf{} & \textbf{MAE}& \textbf{RMSE}& $\mathbf{R}^2$ \\
\midrule
\rdmpnn & 0.0746(0.005) & 0.1040(0.007) & 0.854(0.018) \\
\semg & 0.0424(0.001) & 0.0605(0.002) & 0.951(0.004) \\
\yieldgnn & 0.0408(0.002) & 0.0575(0.002) & \textbf{0.956(0.003)}  \\
\methodwo & \textbf{0.0348(0.002)}& \textbf{0.0544(0.004)} & 0.953(0.009)\\
\bottomrule
\end{tabular}
\begin{tablenotes}
\begin{footnotesize}
\item 
      Each value is the mean and standard deviation (in parentheses), averaging 10 folds. The best performance is highlighted in bold.
      \par
\end{footnotesize}
\end{tablenotes}

\end{threeparttable}

\end{table}

%% file: tables/Pretraining_models_on_BH.tex
\begin{table}[htbp]
\caption{Pre-training-based models performance comparison on the Buchwald-Hartwig dataset}
\label{tbl:Pretraining_models_on_BH}
\centering
\begin{threeparttable}
\begin{tabular}{    
    @{\hspace{12pt}}c@{\hspace{12pt}}
    @{\hspace{10pt}}c@{\hspace{10pt}}
    @{\hspace{10pt}}c@{\hspace{10pt}}
    @{\hspace{10pt}}c@{\hspace{10pt}}
    }
\toprule
\multirow{2}{*}{\textbf{Method}}
\textbf{}&\multicolumn{3}{c}{\textbf{Metrics}} \\
\cmidrule{2-4} 
\textbf{} & \textbf{MAE} & \textbf{RMSE} & $\mathbf{R}^2$ \\
\midrule
\yieldbert & 0.0415(0.001) & 0.0641(0.005) & 0.945(0.008)\\
\chemt & \textbf{0.0311(0.001)} & \textbf{0.0482(0.002)} & \textbf{0.971(0.002)} \\
\methodw & 0.0347(0.001) & 0.0528(0.003) & 0.957(0.006) \\
\bottomrule
\end{tabular}
\begin{tablenotes}
\begin{footnotesize}
\item 
      Each value is the mean and standard deviation (in parentheses), averaging 10 folds. The best performance is highlighted in bold.
      \par
\end{footnotesize}
\end{tablenotes}
\end{threeparttable}
\end{table}

%% file: tables/t_test_yield_ranges.tex
\begin{table}[htbp]
\caption{\methodwo and \chemt performance comparison across different yield ranges on USPTO500MT}
\label{tbl:USPTO_model_comparison}
\begin{threeparttable}
\resizebox{\textwidth}{!}{
\begin{tabular}{
    @{\hspace{8pt}}c@{\hspace{8pt}}
    @{\hspace{4pt}}c@{\hspace{4pt}}
    @{\hspace{4pt}}c@{\hspace{4pt}}
    @{\hspace{4pt}}c@{\hspace{4pt}}
    @{\hspace{4pt}}c@{\hspace{4pt}}
    @{\hspace{4pt}}c@{\hspace{4pt}}
    @{\hspace{4pt}}c@{\hspace{4pt}}
    @{\hspace{4pt}}c@{\hspace{4pt}}
    @{\hspace{4pt}}c@{\hspace{4pt}}
    @{\hspace{4pt}}c@{\hspace{4pt}}
    @{\hspace{4pt}}c@{\hspace{4pt}}
}
\toprule

\textbf{Yield range}                       & \textbf{[0\%,10\%]}          & \textbf{[10\%,20\%]}        & \textbf{[20\%,30\%]}         & \textbf{[30\%,40\%]}        & \textbf{[40\%,50\%]}         & \textbf{[50\%,60\%]}         & \textbf{[60\%,70\%]}         & \textbf{[70\%,80\%]}         & \textbf{[80\%,90\%]}        & \textbf{[90\%,100\%]}      \\
MAE difference (\methodwo - \chemt) & 0.104              & 0.095              & 0.051              & 0.032              & -0.020             & -0.030             & -0.048             & -0.050             & -0.021             & 0.008             \\
T-test p-values                  & 1e-10               & 2e-17               & 2e-1               & 1e-6               & 5e-5               & 3e-12              & 7e-40              & 3e-40              & 6e-8               & 3e-2 \\              

\midrule

\textbf{Cumulative yield range}            & \textbf{[90\%, 100\%]}         & \textbf{[80\%, 100\%]}        & \textbf{[70\%, 100\%]}        & \textbf{[60\%, 100\%]}        & \textbf{[50\%, 100\%]}        & \textbf{[40\%, 100\%]}        & \textbf{[30\%, 100\%]}        & \textbf{[20\%, 100\%]}        & \textbf{[10\%, 100\%]}        & \textbf{[0\%, 100\%]}      \\
MAE difference (\methodwo - \chemt) & 0.008              & -0.006              & -0.018             & -0.024             & -0.025             & -0.024             & -0.020             & -0.015             & -0.011             & -0.009            \\
T-test p-values                  & 3e-2               & 3e-2               & 4e-17               & 3e-37              & 2e-46              & 3e-50              & 4e-35              & 1e-22              & 4e-12               & 1e-8 \\            
\bottomrule
\end{tabular}
}
\end{threeparttable}
\end{table}

%% file: tables/t_test_yield_ranges_CJHIF.tex
\begin{table}[htbp]
\caption{\methodwo and \chemt performance comparison across different yield range on 1,000 sampled non-zero-yielding reactions from CJHIF}
\label{tbl:t_test_yield_ranges_CJHIF}
\centering
\begin{threeparttable}
\resizebox{\textwidth}{!}{%
\begin{tabular}{
    @{\hspace{8pt}}c@{\hspace{8pt}}
    @{\hspace{4pt}}c@{\hspace{4pt}}
    @{\hspace{4pt}}c@{\hspace{4pt}}
    @{\hspace{4pt}}c@{\hspace{4pt}}
    @{\hspace{4pt}}c@{\hspace{4pt}}
    @{\hspace{4pt}}c@{\hspace{4pt}}
    @{\hspace{4pt}}c@{\hspace{4pt}}
    @{\hspace{4pt}}c@{\hspace{4pt}}
    @{\hspace{4pt}}c@{\hspace{4pt}}
    @{\hspace{4pt}}c@{\hspace{4pt}}
    @{\hspace{4pt}}c@{\hspace{4pt}}
    @{\hspace{4pt}}c@{\hspace{4pt}}
}
\toprule
\textbf{Yield range}        & \textbf{(0\%,10\%]}    & \textbf{(10\%,20\%]}        & \textbf{(20\%,30\%]}        & \textbf{(30\%,40\%]}       & \textbf{(40\%,50\%]}        & \textbf{(50\%,60\%]}        & \textbf{(60\%,70\%]}        & \textbf{(70\%,80\%]}        & \textbf{(80\%,90\%]}        & \textbf{(90\%,100\%]}     \\
\midrule
MAE difference (\methodwo - \chemt) &  0.215 & 0.140 & 0.043 &  0.059 & -0.004 & 0.030 & -0.040 & -0.086 & -0.045 & -0.016 \\
T-test p-values   & 2e-1          & 5e-2      & 4e-1               & 3e-1               & 9e-1               & 2e-1              & 8e-3              & 3e-14              & 4e-7               & 4e-2 \\        
\bottomrule     
\end{tabular}
}
\end{threeparttable}
\end{table}

%% file: tables/t_test_yield_ranges_BH.tex
\begin{table}[htbp]
\caption{\methodw and \chemt performance comparison across different yield ranges on the first data split of Buchwald-Hartwig dataset}
\label{tbl:YieldGT_T5Chem_comparison_BH}
\centering
\begin{threeparttable}
\resizebox{\textwidth}{!}{%
\begin{tabular}{
    @{\hspace{8pt}}c@{\hspace{8pt}}
    @{\hspace{4pt}}c@{\hspace{4pt}}
    @{\hspace{4pt}}c@{\hspace{4pt}}
    @{\hspace{4pt}}c@{\hspace{4pt}}
    @{\hspace{4pt}}c@{\hspace{4pt}}
    @{\hspace{4pt}}c@{\hspace{4pt}}
    @{\hspace{4pt}}c@{\hspace{4pt}}
    @{\hspace{4pt}}c@{\hspace{4pt}}
    @{\hspace{4pt}}c@{\hspace{4pt}}
    @{\hspace{4pt}}c@{\hspace{4pt}}
    @{\hspace{4pt}}c@{\hspace{4pt}}
    @{\hspace{4pt}}c@{\hspace{4pt}}
}
\toprule
\textbf{Yield range}                       & \textbf{[0\%,10\%]}    & \textbf{(10\%,20\%]}        & \textbf{(20\%,30\%]}        & \textbf{(30\%,40\%]}       & \textbf{(40\%,50\%]}        & \textbf{(50\%,60\%]}        & \textbf{(60\%,70\%]}        & \textbf{(70\%,80\%]}        & \textbf{(80\%,90\%]}        & \textbf{(90\%,100\%]}     \\
\midrule
MAE difference (\methodw - \chemt) & 0.0027 & 0.0089 & -0.0027 & -0.0039 & 0.0082 & 0.0007 & 0.0073 & 0.0223 & 0.0028 & 0.0030 \\
T-test p-values                  & 6e-2          & 1e-2      & 5e-1               & 3e-1               & 3e-2               & 9e-1              & 2e-1              & 2e-2              & 5e-1               & 6e-1 \\        
\bottomrule     
\end{tabular}
}
\end{threeparttable}
\end{table}

%% file: tables/model_on_each_data_split.tex
\begin{table}[htbp]
\caption{Model performance over each data split on Buchwald-Hartwig}
\label{tbl:model_on_each_data_split}
\centering
\scriptsize{
\begin{threeparttable}
\begin{tabular}{    
    @{\hspace{6pt}}c@{\hspace{6pt}}
    @{\hspace{2pt}}c@{\hspace{2pt}}
    @{\hspace{2pt}}c@{\hspace{2pt}}
    @{\hspace{2pt}}c@{\hspace{2pt}}
    @{\hspace{2pt}}c@{\hspace{2pt}}
    @{\hspace{2pt}}c@{\hspace{2pt}}
    @{\hspace{2pt}}c@{\hspace{2pt}}
    @{\hspace{2pt}}c@{\hspace{2pt}}
    @{\hspace{2pt}}c@{\hspace{2pt}}
    @{\hspace{2pt}}c@{\hspace{2pt}}
    @{\hspace{2pt}}c@{\hspace{2pt}}
    @{\hspace{2pt}}c@{\hspace{2pt}}
    }
\toprule
\multirow{2}{*}{\textbf{Method}}
\textbf{}&\multicolumn{11}{c}{\textbf{Data split}} \\
\cmidrule{2-12} 
\textbf{} & 1 & 2 & 3 & 4 & 5 & 6 & 7 & 8 & 9 &10 & mean(std)\\
\midrule
\yieldbert & 0.0424 & 0.0425 & 0.0408 & 0.0410 & 0.0417 & 0.0411 & 0.0401 & 0.0424 & 0.0403 & 0.0432 & 0.0416(0.001) \\
\chemt & \textbf{0.0323} & \textbf{0.0311} & \textbf{0.0311} & \textbf{0.0314} & \textbf{0.0303} & \textbf{0.0297} & \textbf{0.0315} & \textbf{0.0332} & \textbf{0.0298} & \textbf{0.0309} & \textbf{0.0311(0.001)}\\
\rdmpnn & 0.0758 & 0.0698 & 0.0694 & 0.0758 & 0.0802 & 0.0726 & 0.0727 & 0.0866 & 0.0697 & 0.0734 & 0.0746(0.005)\\
\semg & 0.0440 & 0.0418 & 0.0397 & 0.0432 & 0.0439 & 0.0418 & 0.0433 & 0.0426 & 0.0410 & 0.0424 & 0.0424(0.001)\\
\yieldgnn & 0.0423 & 0.0415 & 0.0391 & 0.0397 & 0.0410 & 0.0418 & 0.0386 & 0.0394 & 0.0406 & 0.0439& 0.0408(0.002)\\
\methodwo & 0.0338 & 0.0339 & 0.0331 & 0.0364 & 0.0360 & 0.0328 & 0.0344 & 0.0379 & 0.0344 & 0.0352 & 0.0348(0.002)\\
\methodw & 0.0363 & 0.0342 & 0.0326 & 0.0371 & 0.0346 & 0.0340 & 0.0338 & 0.0364 & 0.0338 & 0.0343 & 0.0347(0.001)\\
\bottomrule
\end{tabular}
\begin{tablenotes}
\begin{footnotesize}
\item 
      Each value is the model's MAE on the corresponding testing set, the best performance is highlighted in bold. The mean and standard deviation (in parentheses) are obtained by averaging across 10 data splits.
      \par
\end{footnotesize}
\end{tablenotes}
\end{threeparttable}
}
\end{table}

%% file: tables/Ablation_study_module_2.tex
\begin{table}[htbp]
\caption{\methodwo without \MIT performance on the first data split of the Buchwald-Hartwig dataset}
\label{tbl:Ablation_study_module_2}
\centering
\begin{threeparttable}
\begin{tabular}{    
    @{\hspace{8pt}}c@{\hspace{8pt}}
    @{\hspace{8pt}}c@{\hspace{8pt}}
    @{\hspace{8pt}}c@{\hspace{8pt}}
    @{\hspace{8pt}}c@{\hspace{8pt}}
    }
\toprule
\multirow{2}{*}{\textbf{Method}}
\textbf{}&\multicolumn{3}{c}{\textbf{Metrics}} \\
\cmidrule{2-4} 
\textbf{} & \textbf{MAE}& \textbf{RMSE}& $\mathbf{R}^2$ \\
\midrule
\methodwo without \MIT & 0.0490 & 0.0651 & 0.931 \\
\methodwo & \textbf{0.0338} & \textbf{0.0530} & \textbf{0.957} \\
\bottomrule
\end{tabular}
\end{threeparttable}
\end{table}

%% file: tables/hyperparameters_for_USPTO500MT.tex
\begin{table}[htbp]
\centering          
\caption{Hyperparameters and their searched ranges on the USPTO500MT dataset}
\label{tbl:hyperparameters_for_USPTO500MT}
\scriptsize{
\begin{threeparttable}
\begin{tabular}{
    @{\hspace{8pt}}l@{\hspace{8pt}}
    @{\hspace{2pt}}l@{\hspace{2pt}}
    @{\hspace{8pt}}l@{\hspace{8pt}}
    }
    \toprule
    \textbf{Name} & \textbf{Description} & \textbf{Range} \\
    
    \midrule
    eb\_Nlayer & The number of pre-trained self-attention layers to initialize atom features & \underline{0}, \textbf{8} \\
    Nlayer & The number of self-attention layers & 4, \underline{\textbf{5}}, 6 \\
    Nheads & The number of attention heads & \textbf{\underline{16}} \\
    hs & Model dimension & 128, \textbf{\underline{256}}, 512, 1024 \\
    
    \midrule
    e & Epoch & \textbf{\underline{35}} \\
    bs & Batch size & \textbf{\underline{32}} \\
    init\_lr & Initial learning rate &  \underline{\textbf{1e-5}}, 3e-5, 1e-4 \\
    lr\_decay\_step & Learning rate decay patience & 5, \underline{10}, \textbf{15} \\
    lr\_decay\_factor & Learning rate decay factor & 0.85, \underline{0.90}, \textbf{0.95}\\
    dp & Dropout & \textbf{\underline{0}}, 0.1, 0.2 \\
    gnorm & Gradient norm clipping threshold & 0.5, 1, 5, \underline{\textbf{None}} \\
    wd & Weight Decay & \underline{\textbf{0}}, 1e-6, 1e-5 \\
    \bottomrule
    \end{tabular}
\end{threeparttable}
}
\end{table}

%% file: tables/hyperparameters_for_BH.tex
\begin{table}[htbp]
\centering          
\caption{Hyperparameters and their searched ranges on the Buchwald-Hartwig dataset}
\label{tbl:hyperparameters_for_BH}
\scriptsize{
\begin{threeparttable}
\begin{tabular}{
    @{\hspace{8pt}}l@{\hspace{8pt}}
    @{\hspace{2pt}}l@{\hspace{2pt}}
    @{\hspace{8pt}}l@{\hspace{8pt}}
    }
    \toprule
    \textbf{Name} & \textbf{Description} & \textbf{Range}\\
    \midrule
    eb\_Nlayer & The number of pre-trained self-attention layers to initialize atom features & \underline{0}, \textbf{8} \\
    Nlayer & The number of self-attention layers & 1, 2, 3, 4, \textbf{5}, \underline{6}, 7, 8 \\
    Nheads & The number of attention heads & \textbf{\underline{16}} \\
    hs & Model dimension & 128, \textbf{\underline{256}}, 512, 1024 \\
    \midrule
    e & Epoch & \textbf{\underline{300}} \\
    bs & Batch size & \textbf{\underline{32}} \\
    init\_lr & Initial learning rate & 1e-5, \textbf{3e-4}, \underline{1e-4}, 3e-4 \\
    lr\_decay\_step & Learning rate decay patience & 10, \underline{15}, \textbf{20} \\
    lr\_decay\_factor & Learning rate decay factor & \underline{0.85}, \textbf{0.90}, 0.95 \\
    dp & Dropout & \textbf{\underline{0}}, 0.1, 0.2 \\
    gnorm & Gradient norm clipping threshold & \underline{0.5}, 1, 5, \textbf{None} \\
    wd & Weight Decay & \underline{0}, 1e-6, \textbf{1e-5} \\
    \bottomrule
    \end{tabular}
\end{threeparttable}
}
\end{table}

%% file: tables/pretrained_MAT_hyperparameters.tex
\begin{table}[htbp]
\centering           
\caption{Hyperparameters used in the pre-trained MAT model}
\label{tbl:pretrained_MAT_hyperparameters}
\scriptsize{
\begin{threeparttable}
\begin{tabular}{
    @{\hspace{8pt}}l@{\hspace{8pt}}
    @{\hspace{2pt}}l@{\hspace{2pt}}
    @{\hspace{8pt}}l@{\hspace{8pt}}
    }
    \toprule
    \textbf{Name} & \textbf{Description} & \textbf{Value}\\
    \midrule
    eb\_Nlayer & The number of pre-trained MAT self-attention layers & 8\\
    hs\_pretrain & Pre-trained MAT model dimension & 1024\\
    Nheads\_pretrain & Pre-trained MAT attention heads number & 16 \\
    Npff &  The number of dense layers in the position-wise feed-forward block & 1\\
    k & Distance matrix kernel & 'EXP'\\
    dp\_pretrain & Dropout & 0\\
    wd\_pretrain & Weight decay & 0\\
    $\lambda_a, \lambda_d, \lambda_g$ & The scalars weighting the naive self-attention, distance, and adjacency matrices & 0.33\\
    \bottomrule
    \end{tabular}%
\end{threeparttable}
}
\end{table}

%% file: tables/SMILES.tex
\begin{table*}[htbp]
\centering
\tiny
\begin{threeparttable}
\setlength{\tabcolsep}{0pt}%
\begin{tabular}{
    @{\hspace{2pt}}l@{\hspace{2pt}}
    @{\hspace{2pt}}p{0.30\textwidth}@{\hspace{1pt}}
    @{\hspace{1pt}}p{0.26\textwidth}@{\hspace{1pt}}
    @{\hspace{1pt}}p{0.31\textwidth}@{\hspace{1pt}}
}
\toprule
\textbf{Reaction} & \textbf{Reactants} & \textbf{Reagents} & \textbf{Products} \\
\midrule
Reaction 1 of Case 1 in Figure~\ref{uspto_case_study} & COc1ccccc1N1CCNCC1 & O=S(=O)(O)O.[K+].O=[N+]([O-])[O-].O & COc1ccc([N+](=O)[O-])cc1N1CCNCC1 \\
\midrule
Reaction 2 of Case 1 in Figure~\ref{uspto_case_study} & COc1ccccc1N1CCNCC1 & O=S(=O)(O)O.[K+].O=[N+]([O-])[O-].CCOCC.O & COc1ccc([N+](=O)[O-])cc1N1CCNCC1 \\
\midrule
Reaction 1 of Case 2 in Figure~\ref{uspto_case_study} & O=C1CCc2ccccc2O1 & BrBr.S=C=S & O=C1CCc2cc(Br)ccc2O1 \\
\midrule
Reaction 2 of Case 2 in Figure~\ref{uspto_case_study} & COc1ccc(O)cc1 & BrBr.S=C=S & COc1ccc(O)c(Br)c1 \\
\midrule
Reaction 1 of Case 3 in Figure~\ref{uspto_case_study} & COc1ccccc1N=C=O.c1ccc(-c2nsc(N3CCNCC3)n2)cc1 & C1CCOC1.CCN(CC)CC & COc1ccccc1NC(=O)N1CCN(c2nc(-c3ccccc3)ns2)CC1 \\
\midrule
Reaction 2 of Case 3 in Figure~\ref{uspto_case_study} & c1ccc(-c2nsc(N3CCNCC3)n2)cc1.O=C=Nc1ccccc1F & C1CCOC1.CCN(CC)CC & O=C(Nc1ccccc1F)N1CCN(c2nc(-c3ccccc3)ns2)CC1 \\
\midrule
Reaction 1 of Case 4 in Figure~\ref{uspto_case_study} & BrCCBr.COc1cc2c(Nc3ccc(Cl)cc3F)ncnc2cc1O & CN(C)C=O.O=C([O-])[O-].[K+].[K+] & COc1cc2c(Nc3ccc(Cl)cc3F)ncnc2cc1OCCBr \\
\midrule
Reaction 2 of Case 4 in Figure~\ref{uspto_case_study} & BrCCOCCBr.COc1cc2c(Nc3ccc(Cl)cc3F)ncnc2cc1O & CN(C)C=O.O=C([O-])[O-].[K+].[K+] & COc1cc2c(Nc3ccc(Cl)cc3F)ncnc2cc1OCCOCCBr \\
\midrule
\multirow{2}[0]{*}{Reaction A in Figure~\ref{fig: atom_weight_visualization}} & \parbox{0.3\textwidth}{Cc1ccc(cc1)S(Cl)(=O)=O.\\
OCCOCCOCCOCCN1C(=O)c2ccccc2C1=O} & c1ccncc1 & \parbox{0.3\textwidth}{Cc1ccc(cc1)S(=O)(=O)\\
OCCOCCOCCOCCN1C(=O)c2ccccc2C1=O}\\
\midrule
\multirow{2}[0]{*}{Reaction B in Figure~\ref{fig: atom_weight_visualization}} & COc1ccc(cc1){\textbackslash}N=C(/C)c1ccccc1 & 
\parbox{0.3\textwidth}{B(C1=C(C(=C(C(=C1F)F)F)F)\\(C2=C(C(=C(C(=C2F)F)F)F)F)\\
C3=C(C(=C(C(=C3F)F)F)F)FCC1=CC=CC=C1} & COc1ccc(NC(C)c2ccccc2)cc1 \\
\midrule
Reaction C in Figure~\ref{fig: atom_weight_visualization} & c1cc2ccccc2[nH]1.ClS(=O)(=O)c1ccccc1 & C1CCOC1.[Na+].[H-] & O=S(=O)(c1ccccc1)n1ccc2ccccc12 \\
\bottomrule
\end{tabular}
\end{threeparttable}
\caption{SMILES for compounds provided in the manuscripts}
\label{tbl:SMILES}
\end{table*}